\documentclass[12pt]{article}

\usepackage{epsfig}
\usepackage{subfigure}

\newcommand{\half}{\frac 1 2 }
\newcommand{\eg}{{\em e.g.} }
\newcommand{\ie}{{\em i.e.} }
\newcommand{\beq}{\begin{equation}}
\newcommand{\eeq}{\end{equation}}
\newcommand{\be}{\begin{eqnarray}}
\newcommand{\ee}{\end{eqnarray}}

\newcommand{\p}{\partial}

\def\p{\partial}

\def\psidag{\psi^{\dagger}}

\newcommand{\pref}[1]{(\ref{#1})}

\catcode`\@=11
\newcommand{\ccite}[1] {\@ifundefined{b@#1}{\bf ?}{\@nameuse{b@#1}}}
\catcode`\@=12

\begin{document}
\vspace*{1cm}
\centerline{\Large\bf Charge and Statistics of Quantum Hall
Quasi-Particles}
\medskip
\centerline{\large\bf A numerical study of mean values and fluctuations}
\vspace*{-45 mm}
\vskip 55mm
\centerline{\bf H. Kj{\o}nsberg$^{\dagger}$ and J. M. Leinaas}
\medskip
\centerline{Department of Physics, University of Oslo}
\centerline{P.O. Box 1048 Blindern, N-0316 Oslo, Norway}

\vskip 15mm
\centerline{\bf ABSTRACT}
\vspace{.5cm}

We present Monte Carlo studies of charge expectation values and charge 
fluctuations for quasi-particles in the quantum Hall system. We have
studied the Laughlin wave functions for quasi-hole and quasi-electron,
and also Jain's definition of the quasi-electron wave function. The
considered systems consist of from 50 to 200 electrons, and the
filling fraction is $1/3$. For all
quasi-particles our calculations reproduce well the expected values of
charge; $-1/3$ times the electron charge for the quasi-hole, and $1/3$
for the quasi-electron. Regarding fluctuations in the charge, our
results for the quasi-hole and Jain quasi-electron are consistent with
the expected value zero in the bulk of the system, but for the Laughlin
quasi-electron we find small, but significant, deviations from zero
throughout the whole electron droplet. We also present Berry phase
calculations of charge and statistics parameter for the Jain
quasi-electron, calculations which supplement earlier studies 
for the Laughlin
quasi-particles. We find that the statistics parameter
is more well behaved for the Jain quasi-electron than it is for
the Laughlin quasi-electron.

\medskip
\vskip 3mm

\vfil
\noindent
$^{\dagger}$Supported by The Norwegian Research Council.

\eject

\section{Background, motivation and summary of main results}

The perhaps most striking feature of the fractional quantum Hall effect
is the existence of quasi-particles with fractional charge and statistics.
There now exists direct experimental evidence for the existence of
fractionally charged particles \cite{exp}, but much of the evidence for
fractional statistics in the quantum Hall system is still indirect and
based on the theoretical description of the effect.

For the plateau states corresponding to filling fractions $\nu=1/m$, $m$
odd, the many-electron wave functions introduced by Laughlin
\cite{Laughlin} are generally accepted as giving an essentially correct
representation of the true quantum states (under the idealized condition of
a homogeneous background potential). Also the quasi-hole excitations are
described by  simple wave functions, and there are robust arguments for
these to be essentially correct. The charge and statistics of the
quasi-holes have been determined by Berry-phase calculations and agree
with the
claim that they are anyons \cite{Arovas84}. The situation is less clear for
the quasi-electrons.  Different trial wave functions have been suggested,
notably by Laughlin \cite{Laughlin} and  Jain \cite{Jain}, but the 
conclusions
based on these  are less convincing than those for the quasi-holes.

Although the indirect evidence for the charged excitations of the 
fractional
quantum Hall effect to be anyons is rather convincing, a more direct
verification would certainly be interesting. It is a challenge to
establish experimental evidence for the (fractional) statistics of 
the charge
carrying excitations in a similar way as their fractional charge has been
found. However, also on the purely theoretical side it is of interest to
examine further the anyon aspects of the quantum Hall effect.

In a previous paper \cite{Kjonsberg96} we examined the anyon representation
of the Laughlin quasi-holes in some detail. In that paper also the
quasi-electrons were discussed, although much more briefly.
In particular it was
pointed out that in order to reproduce expected values for charge and
statistics parameters by Berry phase calculations, approximations had to be
done which could not readily be justified.  This motivated subsequent
numerical studies of quasi-electrons, by Kj{\o}nsberg and Myrheim
\cite{Kjonsberg98}, for systems with up to 200 electrons. For the charge
parameter bulk values were found which were close to the expected value.
However, for the statistics parameter the rather surprising result was that
no stable bulk value was found. The numerical studies were done by applying
the quasi-electron wave functions introduced by Laughlin. These results
were quite different from the corresponding numerical results obtained for
quasi-holes.

The results obtained in Ref.~\cite{Kjonsberg98} have motivated the
present work, where we examine charge expectation values as well as
charge fluctuations for quasi-holes and quasi-electrons.\footnote{We
are obliged to Hans Hansson for his suggestion to follow up the results
of the Berry phase calculations by examining the charge fluctuations.}
The fractional charge of the physical quasi-particles is expected to be
a sharply defined quantum number, and it is of interest to check
whether this is true for the suggested trial wave functions. There
exist some general arguments which link fractional statistics to
fractional charge \cite{Kivelson85}, and although the precise conditions
for this to be true are not clear, they suggest that the problems
which are seen in Berry phase calculation of the statistics parameter
may also show up in the charge calculations. We have in particular been
interested in examining the possibility of long range fluctuations in
the quasi-electron charge.

The calculations of expectation values and fluctuations of charge have been
done by the Monte Carlo method for quasi-holes and quasi-electrons
corresponding to the $m=3$ state. We have in particular compared results
obtained with Laughlin's \cite{Laughlin} and Jain's \cite{Jain} definitions
of the quasi-electron wave functions. The systems considered have electron
numbers varying from 50 to 200. The main results of the calculations are
the following:

For quasi-holes the expected bulk values of the charge and the charge
fluctuations are well reproduced. The expectation value of the charge is
$-1/3$ (times the electron charge) and the charge fluctuation is zero.
Numerically small statistical fluctuations are present, but no significant
deviations from these values. For Laughlin's quasi-electron the numerical
results for the charge expectation values are consistent with the expected
value $1/3$. However, for the charge fluctuations we find small, but
statistically significant deviations from the value zero in the bulk. This
is the case even for an electron number of 200. For Jain's definition of the
quasi-electron wave function we again reproduce expected results, $1/3$ for
the charge and vanishing fluctuations. Differences between these values and
the numerical results are within small statistical errors.

These results indicate that the problems earlier found in Berry phase
calculations of the statistics parameter \cite{Kjonsberg98} may be an
artifact of Laughlin's special definition and not a signal of long range
effects for a generic quasi-electron state. To check this more directly we
have also examined the charge and statistics parameters found from Berry
phase calculations of the Jain quasi-electron state. We find for
this wave function a stable bulk value of the statistics parameter
consistent with the value $-1/3$.

Throughout the paper, we are using
dimensionless complex coordinates $z=\frac{1}{{\sqrt 2}\ell_B}(x+iy)$,
with $\ell_B= 1/\sqrt{eB}$, and we set $\hbar,c=1$, $B$ being the
magnetic field.

\section{Quasi-particle charge and charge fluctuations}  \label{massedef}

In general it is a subtle problem to define localized and sharp
charges in quantum many body systems. The naive definition of a charge
operator which gives the charge within a finite region $A$ of space is 
(in 2
dimensions)
\beq
{\hat Q}(R) = \int_A {\rm d}^2 x \,  {\hat \rho}(\vec x,\vec x). \label{chop}
\eeq
where ${\hat \rho}(\vec x,\vec y)=\psidag(\vec x)\psi(\vec y)$ is the
single particle density operator and $A$ is taken to be a circular
area of radius $R$. The charge measured relative to the ground state is
defined by a trivial subtraction
\beq
{\hat C}(R) = {\hat Q}(R) - \langle 0 | {\hat Q}(R) |0
\rangle \ \ \ \ \ \label{subt} ,
\eeq
where $|0 \rangle$ is the many body ground state. When the ground state is
represented as a filled Fermi sea, the relative charge ${\hat C}(R)$ can
alternatively be defined by normal ordering the density operator with
respect to the ground state.

States with a well-defined particle number are eigenstates of the total
subtracted charge operator $\hat C(\infty)$; in particular
$\hat C(\infty)|0\rangle =0$. For the charge operator corresponding to a
finite area $A$, this is not the case, even if the radius $R$ is taken to
be very large. This can be seen by  considering the charge
fluctuation\footnote{ Clearly, the subtraction in
\pref{subt} does not change the  fluctuations, \ie $(\Delta C(R))^{2}
= (\Delta Q(R))^{2}$. }
\beq
(\Delta Q(R))^{2} = \langle {\hat Q}(R)^2 \rangle -
\langle {\hat Q}(R) \rangle ^2 \;,    \label{odisp}
\eeq
which does not vanish even for the ground state. In a relativistic field
theory the charge fluctuation in fact diverges due to contributions from
particle-antiparticle pairs of arbitrary high momenta. In a
non-relativistic theory the fluctuation is finite, due to the finite depth
of the Fermi sea, although it will in general be large. The case we
consider here is even more well behaved, since the particles are restricted
to the lowest Landau level. A deep Fermi sea would correspond to a
situation with many filled Landau levels.

For states with short range correlations, we  expect the fluctuation
to be an edge effect, and to demonstrate this more explicitly we rewrite
the charge fluctuation in the following form,
\beq
(\Delta Q(R))^{2} = - \int_A {\rm d}^{2} x \, \int_{A^{C}} {\rm d}^{2} y \,
c(\vec x,\vec y)      \label{disp}
\eeq
where $A^{C}$, is the complement of the area $A$, and $c(\vec x,\vec y)$
is the density-density correlation function defined by,
\beq
c(\vec x,\vec y) = \langle \hat\rho (\vec x) \hat\rho (\vec
y)\rangle - \langle \hat\rho (\vec x) \rangle \langle \hat\rho (\vec
y) \rangle \;.
\label{paircorr}
\eeq
For a homogeneous ground state the correlation function only depends 
on the
relative distance, $c(|\vec x -\vec y|) =c(r)$, and with an exponential
fall off for large $r$  the integrals in Eq.~\pref{disp}
only get contributions close to the boundary of $A$ and $A^C$. This
gives for the charge fluctuation a linear dependence on the radius $R$,
when $R$ is much larger than the correlation length.

\begin{figure}[htb]
\begin{center}
{\psfig{figure=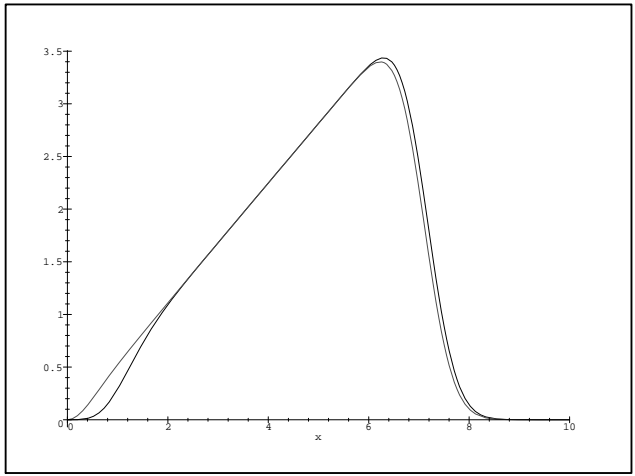,
angle=270,width=7.5cm}}
\end{center}
\caption[]{\footnotesize The charge fluctuation $(\Delta Q(R))^2$
for a system with 50 electrons and filling factor $m=1$. The two curves show
the fluctuations of the ground state and of the state for one quasi-hole. }
\label{fig1a}
\end{figure}

For the incompressible states of the quantum Hall system, the density
correlations are short range, with a correlation length of the order of
the magnetic length $\ell_B$.
To illustrate the form of the charge fluctuation $(\Delta Q(R))^{2}$ in 
this
case, we consider the case of a filled lowest Landau level, and
from now on all lengths are measured in units of $\sqrt 2 \ell_B$ and
integrals are given on dimensionless form.
In the limit
of an infinite system we then have an analytic expression for the
correlation function,
\beq
c(r) = \frac{1}{\pi}\left(\delta(\vec r) -
\frac{1}{\pi}\,e^{-r^{2}}\right) \; ,
\label{expfall}
\eeq
and the charge fluctuation  can easily be calculated.
This is the case even for a finite system, and in Fig.~\ref{fig1a} the
functional form of $(\Delta Q(R))^{2}$ is shown for an electron droplet
corresponding to 50 electrons. The linear dependence is clearly seen 
for an
intermediate range, where $R$ is larger than $\ell_B$ but smaller  than the
size of the system. For the Laughlin states we expect a similar
behaviour. In this case we can use Laughlin's plasma analogy to demonstrate
the exponential fall-off of the correlation function, a behaviour
corresponding to the screening of charges in the plasma. Earlier
numerical calculations of the correlation function \cite{laughlingp}
also show the
exponential fall-off, and we have by direct numerical
calculations found a functional form of $(\Delta Q(R))^{2}$ for the
$\nu = 1/3$ state which is similar to the one shown in
Fig.~\ref{fig1a}.

It is a non-trivial problem to construct a local charge operator which is
insensitive to the background fluctuations of the ground state. In the
context of a one-dimensional fermi system with fractionally charged
solitons, the problem was solved by Kivelson and Schrieffer
\cite{Kivelson82} and  Bell and Rajaraman \cite{Rajaraman82} who defined
a charge operator using a smooth spatial cutoff.
However, in 2 (and higher) dimensions this is not sufficient, also a
cutoff in energy is needed, as discussed by Goldhaber
and Kivelson \cite{Goldhaber91}. In principle such a definition could be
used also for the fractional quantum Hall states, however to implement
this in the form of an explicitly defined charge operator seems difficult.

In this paper we have not made any attempt to introduce such a redefined
charge operator. Instead we have used, as the relevant measure of the
sharpness of a localized charge, the subtracted quantity
\beq
D(R) \equiv  \Delta Q(R) - \Delta Q_0(R) \ \ \ \ \ ,
\label{qfl}
\eeq
where $\Delta Q_0(R)$ is the fluctuation of the ground state. Each of the
two terms in this expression are finite, and the difference is therefore
well defined. The use of this expression is based on the intuitive idea
that if an excited state can be characterized by a sharply defined,
localized charge, that means that the charge fluctuation rapidly
approaches the ground state value when the integration area $A$ is
extended outside the region which characterizes the size of the excitation.
In fact, it seems reasonable to assume that if the fluctuation vanishes
for any reasonably defined local charge operator, then also the simple
difference \pref{qfl} will vanish. The only problem with the approach used
here is of numerical character. The fluctuation will be a  small number
calculated as the difference between large numbers, and this put strong
limits on the accuracy of the numerical calculations.

\begin{figure}[htb]
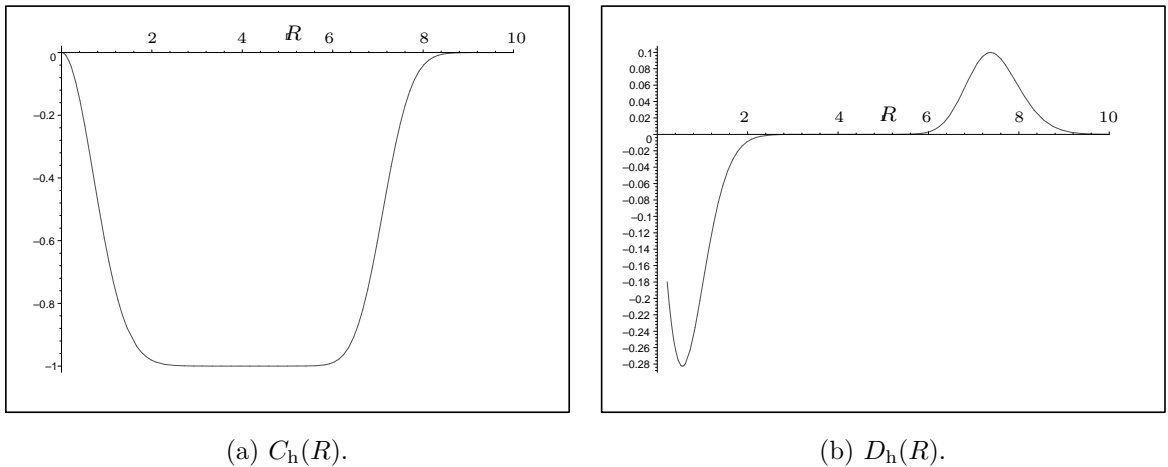

\begin{center}
\begin{tabular}{cc}
\subfigure[$C_{\rm
h}(R)$.]{\psfig{figure=f1m1ladn.ps,angle=270,width=7.5cm}} &
\subfigure[$D_{\rm
h}(R)$.]{\psfig{figure=f2m1flukt.ps,angle=270,width=7.5cm}}
\end{tabular}
\put(-391,79){$_{_2}$}
\put(-357,79){$_{_4}$}
\put(-324,79){$_{_6}$}
\put(-288,79){$_{_8}$}
\put(-255,79){$_{_{10}}$}
\put(-167,49){$_{_2}$}
\put(-131,49){$_{_4}$}
\put(-98,49){$_{_6}$}
\put(-63,49){$_{_8}$}
\put(-31,49){$_{_{10}}$}
\put(-114,48){${_R}$}
\put(-339,79){${_R}$}
\end{center}
\caption[]{\footnotesize Quasi-hole charge $C_{\rm h}(R)$ and charge
fluctuation $D_{\rm h}(R)$ for a system with 50 electrons and filling
fraction $1/m=1$.}
\label{fig1}
\end{figure}

For the sake of illustration we show in Fig.~\ref{fig1} the charge 
expectation
value and the charge fluctuation, with such a subtraction included, for the
special case of a Laughlin quasi-hole at the integer filling factor $1$. In
this case all the quantities needed in order to calculate the 
functions $C(R)$
and $D(R)$ can be found analytically\footnote{The wave functions will 
be given
in full detail in section \ref{s3}, but for now our point is 
only that of
illustration.}, and we show the results for a system of 50 
electrons. The
excess charge, here negative, exponentially  builds up to the 
plateau value
$(-1)$ as $R$ increases from zero, and the charge fluctuations 
vanish once the
plateau is reached. We also see edge effects at the droplet 
boundary $R\approx
7.1$. The figures clearly show that our
definitions  in Eqs.~(\ref{subt}) and (\ref{qfl}) reveal the basic
properties of interest.

Before concluding this section on general definitions and relations,
we shall introduce a quantity that will be very useful in the
coming calculations. Consider
\beq
{\hat F}(R) = \int_A {\rm d}^2 \!u {\rm d}^2 \!v{\hat \rho}_2(u,v),
\eeq
where ${\hat \rho}_2(u,v)$ is the diagonal two-electron density
operator. Using
\beq
{\hat \rho}(u){\hat \rho}(v) =  {\hat \rho}_2(u,v) +
\delta(u-v){\hat \rho}(u,v)
\eeq
 we can then express the fluctuation of the operator ${\hat Q}$ as
\be
(\Delta Q(R))^2 &=& \langle{\hat Q}^2(R)\rangle - \langle{\hat
Q}(R)\rangle^2
\label{nofl} \\
&=& Q(R) + F(R) - (Q(R))^2.  \label{qf}
\ee
The functions $Q,F$ are expectation values of ${\hat Q},{\hat F}$ in
the specific state. In terms of the $N$-electron normalized wave
function
for the state
they are
\beq
Q(R) = N \int_A d^2\! u \int d^2\!z_2 \cdots
d^2\!z_N \mid \psi(u,z_2,\cdots ,z_N) \mid^2  \label{Qf}
\eeq
\beq
F(R) = N(N-1) \int_A d^2\! u d^2\! v  \int d^2\!z_3 \cdots
d^2\!z_N \mid \psi(u,v,z_3,\cdots ,z_N) \mid^2     \label{Ff}
\eeq
The expression (\ref{qf}) will be used when investigating the charge
fluctuations of the Laughlin quasi-electrons, 
whereas (\ref{nofl}) will be
used in the case of Laughlin's ground state and quasi-hole 
state as well as
for Jain's quasi-electrons.

\section{Numerical methods to compute $Q$ and $\Delta Q$}     \label{s3}
\subsection{General introduction}     \label{s31}

In all our computations we have used Monte Carlo integration with
importance sampling according to the Metropolis algorithm
\cite{james, laughlingp}, and a brief discussion of some points of
special importance for our calculations are given in appendix
\ref{numerikk}. The
specific probability distribution used
to generate electron
configurations varied from case to case. Each of them will be given in
the sections following below, where we give descriptions of our
methods that are detailed enough to allow the calculations to be
reproduced.

Notice that in order to find the quasi-particle charge fluctuation
$D(R)$ we need to find the two numbers
$\Delta Q(R)$ and
$\Delta Q_0(R)$, and then the difference between them. In general,
this difference is very small compared to the numbers themselves.
Since
our aim is to decide if the difference is
significantly different from zero, both $\Delta
Q(R)$ and $\Delta Q_0(R)$ must be known with great precision.
This
is itself a challenging problem since each of them in turn is found
from a
difference between the numbers $\langle {\hat Q}^2(R)\rangle$ and
$\langle {\hat Q}(R)\rangle^2$ which are large for large $R$. For
all examined cases except for
Laughlin's quasi-electrons, this problem has been solved by evaluating
$\langle {\hat Q}^2\rangle$ and $\langle {\hat Q}\rangle^2$
simultaneously, and benefit from the fact that the statistical errors
then are correlated and tend
to cancel.
For the Laughlin quasi-electrons this method is
inappropriate. Instead we have found values for $Q$ and $F$ defined in
Eqs.~(\ref{Qf}) and (\ref{Ff}).
In the following discussion we will use the subscripts ``h, Le"
and ``Je" to refer to quasi-holes, Laughlin quasi-electron and
Jain quasi-electrons, respectively. The subscript ``$0$" refers
to the ground state.

\subsection{The Laughlin ground state and quasi-hole state} \label{s32}

The definitions (\ref{subt}) and (\ref{qfl}) of the charge and
charge fluctuations for the quasi-particle, require knowledge of
the mean value and fluctuations of the ground state charge. For
$N$ electrons at filling
fraction $1/m$ the ground state is given by Laughlin's wave
function
\beq
\psi_0(z_1,\cdots z_N^*)  =  e^{-\half \sum_{i=1}^N\mid z_i \mid^2 }
\prod_{k <l} ( z_k - z_l)^m, \label{laugr}
\eeq
This  describes an electron droplet with uniform particle
density (in the bulk) and a radius
approximately $\sqrt{mN}$ \cite{laughlingp, halpmorf}.
In this case the numerical computations are simple.
We use the true electron density to generate the electron
configurations $(z_{1\alpha},\cdots,z_{N\alpha})$, that is we generate
events according to the probability density,
\beq
p_0(z_1,\cdots z_N^*) = \frac{1}{I_0}|\psi_0(z_1,\cdots z_N^*)|^2,
\label{prfir}
\eeq
where
\beq
I_0 = \int {\rm d}^{2N}\!z
\,|\psi_0(z_1,\cdots z_N^*)|^2
\eeq
is the normalization integral. For each specific configuration
$\alpha$ the charge $Q_{\alpha}$ inside the area $A=\pi R^2$ is found
simply by counting the $z$:s that satisfy $|z|\le R$.
From $n$ configurations the expectation values of ${\hat Q}(R)$ and
${\hat Q}^2(R)$ are estimated (simultaneously)  by
\beq
\langle 0 | {\hat Q}(R)|0 \rangle = \frac{1}{n} \sum_{\alpha = 1}^{n}
Q_{\alpha},  \hspace{1cm} \langle 0 | {\hat Q}^2(R)|0 \rangle =
\frac{1}{n} \sum_{\alpha = 1}^{n} Q_{\alpha}^2,  \label{simmeth}
\eeq
and the ground state fluctuation $\Delta Q_0(R)$ is then found using
Eq.~(\ref{nofl}).

The same method can also be used for a Laughlin quasi-hole, which is given
by the wave function
\beq
\psi_{\rm h}(z_1,\cdots z_N^*)  =  e^{-\half \sum_{i=1}^N\mid z_i
\mid^2 }
\prod_{k <l} ( z_k - z_l)^m \prod_{i=1} ( z_i - z_0)\ \ \ \ \ .  
\label{lauhu}
\eeq
This expression describes an electron droplet with an electron density
which is the same as in the ground state, except for a small 
depleted region
around the point $z_0$, the position of the quasi-hole
\cite{laughlingp}. Again we choose the probability distribution used
to generate electron configurations
to be the (normalized) electron probability density itself;
\beq
p_{\rm h}(z_1,\cdots z_N^*) = \frac{1}{I_{\rm h}}|\psi_{\rm
h}(z_1,\cdots z_N^*)|^2,
\hspace{5mm} {\rm where} \hspace{3mm}I_{\rm h} = \int {\rm d}^{2N}\!z
\,|\psi_{\rm h}(z_1,\cdots z_N^*)|^2 .
\eeq
For $z_0=0$ the expectation values $\langle {\rm h} | {\hat
Q}(R)|{\rm h}
\rangle $ and $\langle {\rm h} | {\hat Q}^2(R)|{\rm h}
\rangle$ are then estimated, using the procedure
described above.

\subsection{The Laughlin quasi-electron state}      \label{s33}

The simple numerical method described above can not be used in the
case of a Laughlin quasi-electron. The reason for this can immediately
be seen from the wave function, which reads
\beq
\psi_{\rm Le}(z_1,\cdots z_N^*) = e^{-\half \sum_{j=1}^N |z_j |} 
\prod_{i=1}^N
(\p_{z_i}-z_0^*)
\prod_{k<l}(z_k - z_l)^m. \label{helv}
\eeq
In this case the wave function contains derivatives, and these must
be evaluated analytically before the expressions can be
used in a computer program.
Although in principle straightforward, we know of no efficient way of
doing this for sufficiently many electrons\footnote{
That the Laughlin quasi-electron represents challenging numerical
problems is  well known. Several authors have
calculated the single electron density with a quasi-electron
placed at the origin of a circular droplet, and with
results that are not in total agreement with one another. The
discrepancies are restricted to the behaviour close to the
origin. Haldane and Rezayi \cite{hr}, and later Morf and Halperin
\cite{halpmorf}, found, as opposed to Laughlin
\cite{laughlingp}, that the single electron density has a dip at the
origin. In Ref.~\cite{halpmorf} this dip even drops below $1$.   }.

To avoid this problem, we  have considered two
different methods for computing the charge and
charge fluctuations of a Laughlin quasi-electron. The first
uses ``brute force'', and the numerical convergence is slow,
although much better for small than for large $R$. The second method
converges faster, and is best fitted for the bulk. Hence, the domains
of validity of the two
methods are complementary, which allows us to calculate for the whole
range of $R$. Also there is some overlap where we can check against one
another the results obtained by use of the two methods.

\subsubsection*{Method 1}

We will here present the method most appropriate for small
$R$. Consider then the expectation value of the charge operator when
the quasi-electron is put at the origin;
\be
&&  Q_{\rm Le}(R) = \langle {\rm Le} |{\hat Q}(R)|{\rm Le} \rangle \\
&=& \frac{N}{I_{\rm Le}}
\int_A {\rm d}^2\! z_1 \int {\rm d}^{2(N-1)}\! z \, |\psi_{\rm
Le}(z_1,\cdots, z_N^*)|^2      \label{qno} \\
&=&  \frac{N}{I_{\rm Le}} \int_A {\rm d}^2\! z_1 \int {\rm
d}^{2(N-1)}\! z \,
 e^{-\sum_{i=1}^N \mid z_i \mid^2 }
\prod_{k <l} \mid z_k - z_l\mid ^{2m} \prod_{j=2}^N (\mid z_j \mid^2
-1)  \mid \sum_{i=2}^N \frac{m}{z_1 - z_i} \mid^2
\label{rew}\\
&=&  \frac{N}{I_{\rm Le}} \int_A {\rm d}^2\! z_1 \int {\rm
d}^{2(N-1)}\! z \, p^{\prime}_{\rm Le}(z_1,\cdots,z_N^*)
\label{fakp}
\ee
where $I_{\rm Le}$ is the normalization integral, and
$p^{\prime}_{\rm Le}$
is the integrand in Eq.~(\ref{rew}).
The expression in Eq.~(\ref{rew}) is obtained by partially integrating
the coordinates $z_2,\cdots,z_N$, but differentiating the VanderMonde
determinant directly with respect to $z_1$,
which is integrated over the finite area $A$. We emphasize that
$p^{\prime}_{\rm Le}$ is not the true electron density of the 
system, in
fact, since it takes on negative as well as positive value, it cannot
be taken as a probability density. Nevertheless, the
normalization integral $I_{\rm Le}$ can be expressed in terms of
$p^{\prime}_{\rm Le}$,
\beq
I_{\rm Le} =  \int {\rm d}^{2N}\! z \, |\psi_{\rm Le}(z_1,\cdots,
z_N^*)|^2 = \int {\rm d}^{2N}\! z \, p^{\prime}_{\rm
Le}(z_1,\cdots,z_N^*) \ \ \ \ \ ,
\eeq
and this enables us to write,
\beq
Q_{\rm Le}(R)
= N \frac{\int_A {\rm d}^2\! z_1 \int {\rm
d}^{2(N-1)}\! z \, p^{\prime}_{\rm Le}(z,z^*) }{\int {\rm
d}^{2N}\! z \, p^{\prime}_{\rm Le}(z,z^*)  }
= N \frac{\int_A {\rm d}^2\! z_1 \int {\rm
d}^{2(N-1)}\! z \, |p^{\prime}_{\rm Le}(z,z^*)| \,
{\rm sgn}(p^{\prime}_{\rm Le})
}{\int {\rm
d}^{2N}\! z \, |p^{\prime}_{\rm Le}(z,z^*)| \,
{\rm sgn}(p^{\prime}_{\rm
Le})} \ \ \ ,  \label{rats}
\eeq
where ${\rm sgn}(p^{\prime}_{\rm Le})$ is the sign function.
Eq.~(\ref{rats}) tells us that if the electron coordinates 
are generated
according to the absolute value $|p^{\prime}_{\rm Le}(z,z^*)|$ then
the expectation value $Q_{\rm Le}(R)$ can be estimated as the
ratio between the Monte Carlo expectation values of
$S(R-|z_1|)\,{\rm sgn}(p^{\prime}_{\rm Le})$ and ${\rm sgn}
(p^{\prime}_{\rm Le})$, where $S(x)$ is $1$ when $x\geq
0$ and $0$ otherwise. In practice we thus do
as follows: For each electron configuration $\alpha$ determine whether
particle $1$ is inside $R$. If it is, set $t_{\alpha}=\pm 1$
according to the sign of $p^{\prime}_{\rm Le}(z,z^*)$. Otherwise set
$t_{\alpha}=0$. Independent of the position of $z_1$ set $n_{\alpha} =
\pm 1$ according to the sign of $p^{\prime}_{\rm Le}$.
The expectation value is then found by
\beq
Q_{\rm Le}(R) =  \frac{\sum_{\alpha=1}^n t_{\alpha}}{\sum_{\alpha=1}^n
n_{\alpha}}.
\eeq
Notice that the function
$|p^{\prime}_{\rm Le}(z,z^*)|$ is not
symmetric in all variables since it treats $z_1$ in a special
way. This implies that the numerical convergence rate is $N$ 
times slower
than it would have been if we could use a symmetric function,
everything else  being equal.

Since the electron coordinates  were not generated according to the
true electron density, the expectation
value $\langle {\hat Q}^2(R)\rangle$ can not be
found using the simple method described for the Laughlin ground state
and quasi-hole. Instead we turn to Eq.~(\ref{qf}) and compute
\be
&& F_{\rm Le}(R) = \langle {\rm Le} |{\hat F}(R)|{\rm Le} \rangle \\
&=& \frac{N(N-1)}{I_{\rm Le}}
\int_A {\rm d}^2\! z_1 {\rm d}^2\! z_2 \int {\rm d}^{2(N-2)}\! z \,
|\psi_{\rm
Le}(z_1,\cdots, z_N^*)|^2        \label{fno}   \\
&=& \frac{N(N-1)}{I_{\rm Le}}  \int_A {\rm d}^2\! z_1 {\rm d}^2\! z_2
\int {\rm d}^{2(N-2)}\! z \,     e^{- \sum_{i=1}^N |z_i|^2 }
\prod_{k <l} |z_k - z_l| ^{2m} \prod_{k=3}^N (|z_k|^2-1)  \\
&& \hspace{4cm} \times   |\sum_{i\neq 1}^N \frac{m}{z_1 - z_i}
\sum_{j\neq 2}^N
\frac{m}{z_2 - z_j} + \frac{m}{(z_1 - z_2)^2}|^2 .  \label{stygg}
\ee
The VanderMonde determinant has now been differentiated with respect
to both $z_1$ and $z_2$, while the other coordinates were
integrated by parts. The integrand in Eq.~(\ref{stygg}) is more
complicated than
$p^{\prime}_{\rm Le}$ in Eq.~(\ref{fakp}). However, if we now define
$p^{\prime}_{\rm Le}$ to be the integrand of Eq.~(\ref{stygg}), then
the normalization
integral $I_{\rm Le}$ can still be expressed as $I_{\rm Le} = \int
{\rm d}^{2N}\! z \, p^{\prime}_{\rm Le}(z,z^*)$, and the value of $F_{\rm
Le}(R)$ be estimated as the ratio between two Monte Carlo estimates,
analogously to the method above.
Notice that in the present case the function $|p^{\prime}_{\rm
Le}(z,z^*)|$ treats two coordinates differently from the others. The
convergence rate of this non-symmetric treatment will be $N(N-1)$ times
slower than a symmetric one, again everything else being equal.

The slow convergences for large $N$ that we have referred to, imply
problems for the numerical calculations discussed here.
Nevertheless, for small values of $R$ we do obtain well converged
results. These can be used as a complement to the bulk results
found by use of the method of the next section. In addition, the ranges of
validity of the two methods do have some overlap, and we hence check
the results against one another. We also hope that the treatment
of ``negative probability densities'' is instructive, and would like
to mention that also Ref.~\cite{halpmorf} contains a discussion of this
topic. It will be used again in the next section.

\subsubsection*{Method 2}

This section presents a method that improves
Method 1 in two important ways: First, a function symmetric in
all electron coordinates is used to generate configurations, and
thus we achieve a huge
improvement in the convergence rate.
Second, the same electron configurations are now used for calculating
both
$Q_{\rm Le}(R)$ and $F_{\rm Le}(R)$.
However, to determine the functions
we need to use numerical derivatives in addition to the Monte Carlo
integration.

Define the quantity
\beq
{\tilde p}_{\rm Le}(z_1,\cdots,z_N^*) = e^{-\sum_{i=1}^N | z_i|^2}
\prod_{j=1}^N (| z_j|^2 -1) \prod_{k<l}| z_k - z_l|^{2m}.
\label{fakden}
\eeq
Of course, ${\tilde p}_{\rm Le}$ is nothing else than the integrand
appearing
if we use integration by parts to {\em all} coordinates in
Eqs. (\ref{qno}, \ref{fno}), and it is clear that
\beq
I_{\rm Le} =  \int {\rm d}^{2N}\! z \, |\psi_{\rm Le}(z_1,\cdots,
z_N^*)|^2 = \int {\rm d}^{2N}\! z \, {\tilde p}_{\rm
Le}(z_1,\cdots,z_N^*).
\eeq
In addition, define the auxiliary functions,
\beq
{\widetilde Q}(R) = \frac{N}{I_{\rm Le}} \int_A {\rm d}^2 z_1 \int {\rm
d}^{2(N-1)} z \,{\tilde p}_{\rm Le}(z_1,\cdots,z_N^*)  ,
\label{tilQ}
\eeq
\beq
{\widetilde F}(R_1,R_2) = \frac{N(N-1)}{I_{\rm Le}} \int_{A_1} 
{\rm d}^2 z_1
\int_{A_2} {\rm d}^2 z_2 \int {\rm d}^{2(N-2)} z \,  {\tilde
p}_{\rm Le}(z_1,\cdots,z_N^*) .  \label{tilF}
\eeq
Notice that the areas $A_1$ and $A_2$ in general are different.

The auxiliary functions ${\widetilde Q}(R)$ 
and ${\widetilde F}(R_1,R_2)$
are closely related to the functions, $Q_{\rm Le}(R)$ 
and $ F_{\rm Le}(R)$
that we want to compute. When partial integration is applied 
to the
variable $z_1$, boundary terms will appear since the integration area
$A$ covers only a part of the full droplet. The boundary terms can be
expressed in terms of ${\widetilde Q}(R)$ and 
${\widetilde F}(R_1,R_2)$ and
their first two derivatives and give the relations
\be
Q_{\rm Le}(R) &=& {\widetilde Q}(R) + c_1(R) \frac{{\rm d} 
{\widetilde
Q}(R)}{{\rm d}R} + c_2(R)
\frac{{\rm d}^2{\widetilde Q}(R)}{{\rm d}R^2}, \label{tilq} \\
F_{\rm Le}(R) &=& {\widetilde F}(R,R) + 
c_1(R)\left( \frac{\p {\widetilde
F}}{\p R_1} + \frac{\p
{\widetilde F}}{\p R_2}\right)_R \nonumber \\
&& + c_2(R)\left( \frac{\p^2 {\widetilde F}}{\p R_1^2} + 
\frac{\p^2
{\widetilde F}}{\p R_2^2}\right)_R + c_1^2(R) \left( \frac{\p^2
{\widetilde F}}{\p R_1 R_2 }\right)_R \nonumber \\
&& + c_1(R) c_2(R) \left( \frac{\p^3 {\widetilde F}}{\p R_1^2 R_2} 
+ \frac{\p^3
{\widetilde F}}{\p R_1 R_2^2}\right)_R + c_2^2(R) \left( \frac{\p^4
{\widetilde F}}{\p R_1^2 R_2^2}\right)_R. \label{tilf}
\ee
The subscript $R$ means that both $R_1$ and $R_2$ should be set equal to
$R$ after differentiating. Eqs.~(\ref{tilq}) and (\ref{tilf}) are exact
equalities, and the derivations are shown in appendix \ref{relst}. We
have used the notation
\be
c_1(R) = \frac{4R^4-7R^2+1}{4R(R^2-1)^2},  &&
c_2(R) = \frac{1}{4(R^2-1)}.  \label{c1c2}
\ee

The numerical method for obtaining $Q_{\rm Le}(R)$ and 
$F_{\rm Le}(R)$
is as follows: First generate coordinates according to the
absolute value $|{\tilde p}_{\rm Le}|$,
and estimate ${\widetilde Q}(R)$
and ${\widetilde F}(R_1,R_2)$ simultaneously in the way described
above, but now with the
advantage of having a generating function that is symmetric in the
electron coordinates. The computation is done for
$R,R_1,R_2=hk$ where $h$ is a fixed grid spacing and $k=0,1,\cdots,K$.
The maximum value $K$ is taken large enough that the radii take values
larger than the radius of the electron droplet, which for $N$
electrons at filling fraction $1/m$ is $\sqrt{mN}$. From the resulting
one- and two-dimensional grid of data  we estimate the derivatives by
the formula
\beq
\frac{{\rm d}{\widetilde Q}(R)}{{\rm d} R} \approx 
\frac{1}{2h}({\widetilde
Q}(R+h)-
{\widetilde Q}(R-h))    \label{dert}
\eeq
with similar expressions for higher derivatives. The values 
for $Q_{\rm
Le}(R)$ and $F_{\rm Le}(R)$ are found by use of Eqs.~(\ref{tilq}) and
(\ref{tilf}), and the value of $\Delta Q_{\rm Le}(R)$ by use of
Eq.~(\ref{qf}). In the computation we have used the two step sizes
$h=0.1$ and
$h=0.2$, and as will be discussed in detail in section 4, we have
sufficient control over the errors introduced by
the discrete differentiation.

\subsection{The Jain quasi-electron}       \label{s34}

We have also computed the charge and charge fluctuations  for the
wave function defined by Jain.
The wave function has the form
\cite{Jain}
\be
\psi_{\rm Je} &=& {\cal P}e^{-\half\sum_{i=1}^N|z_i|^2} \left|
\begin{array}{cccc}
                                  z_1^* & z_2^* &\cdots & z_N^* \\
                                     1  & 1  &  \cdots &1     \\
                                    z_1 & z_2 & \cdots & z_N  \\
                                   \vdots & \vdots & \vdots & \vdots
\\
                                  z_1^{N-2} & z_2^{N-2} & \cdots &
z_N^{N-2}
                                          \end{array} \right|
                \prod_{k<l}(z_k - z_l)^{m-1} \label{detwave}\\
&=& {\cal P} e^{-\half\sum_{i=1}^N| z_i|^2} \prod_{k<l}(z_k -
z_l)^{m-1}
\sum_{j=1}^N \left( z_j^* (-1)^j \prod_{k<l;k,l\neq j}(z_k - z_l)
\right)\\
&=& {\cal P} e^{-\half\sum_{i=1}^N| z_i|^2} \sum_{j=1}^N \left(
z_j^* \frac{1}{\prod_{i\neq j} (z_j - z_i)}  \right) \prod_{k<l}(z_k -
z_l)^{m}.   \label{jainq}
\ee
Here ${\cal P}$ means projection onto the lowest Landau level. The
importance of this projection has been studied \cite{japim}, and it
turns
out that even without ${\cal P}$ most of the wave function is already
in the lowest Landau level. Our calculations, which are performed for
the unprojected as well as the projected state, give results that are
                                     in accordance with this.

Following the general scheme for projection onto the lowest Landau
level \cite{girvjach}, we let $z_j^* \rightarrow \p_{z_j}$ with the
differentiation operator acting on everything except the exponential
factor $e^{-\half\sum_{i=1}|z_i|^2}$. This yields
\beq
\psi_{\rm Je} = e^{-\half\sum_{i=1}^N| z_i|^2} \prod_{k<l}(z_k -
z_l)^{m}
\sum_{j=1}^N \frac{1}{\prod_{i\neq j} (z_j - z_i)} \left( -\sum_{k\neq
j} \frac{1}{(z_j - z_k)} + \frac{m}{ \sum_{k\neq j}(z_j - z_k)}
\right).
\label{pjain}
\eeq
This expression is symmetric in all coordinates. In addition, the
squared absolute value is the true electron density in the
system, so in order to find
$Q_{\rm Je}(R)$ and $\Delta Q_{\rm Je}(R)$ we can simply adopt the
method used
for the Laughlin ground state and quasi-hole state,
Eq.~(\ref{simmeth}), now with
$p_{\rm Je}=|\psi_{\rm Je}|^2/I_{\rm Je}$ as the probability
distribution used in
the Metropolis algorithm.
For the unprojected state the method is of course similar, and the
expression in Eq.~(\ref{jainq}), now without the projection operator
${\cal P}$, is used for $p_{\rm Je}$.

\section{Results of charge mean value and charge fluctuation
computations}

This section presents the results following from the calculations
described in the previous section. For $m=3$, \ie for filling fraction
$1/3$ of the lowest Landau level, we have calculated charge and charge
fluctuations in the cases of a Laughlin quasi-hole, a Laughlin
quasi-electron and a Jain quasi-electron. In all three cases the
quasi-particle has been located at the center of a circular electron
droplet. For the quasi-hole we have made calculations for electron
droplets consisting of 50 and 100 particles. Systems with 50, 100 and
200 electrons were considered in the case of a Laughlin quasi-electron,
whereas we did calculations for 50 electrons with the Jain
quasi-electron in the system. We find that for all quasi-particles the
numerical calculations reproduce well the expected bulk values of the
charge mean values. There are finite-size effects, due to the limited
number of electrons, which are most dominant for 50 electrons in the
case of the Laughlin quasi-electron. However, these effects
are constrained to the regions close to the position of the
quasi-particle and to the edge of the electron droplet. For
$100$ and $200$ electrons (and even for $50$ electrons in the case of a
quasi-hole or a Jain quasi-particle) the bulk values, $-1/3$ for
quasi-holes and $+1/3$ for quasi-electrons, are reproduced within small
statistical errors. Charge is then measured in units of the electron
charge. For the quasi-hole and Jain quasi-electron the calculations
reproduce, again within small statistical errors, the expected bulk
value zero for the charge fluctuations. However, for the Laughlin
quasi-electron we find  larger fluctuations than in the other two cases.
At an absolute scale they are small, but they are significantly
different from zero within the small errors of the calculation.

\begin{figure}[htb]
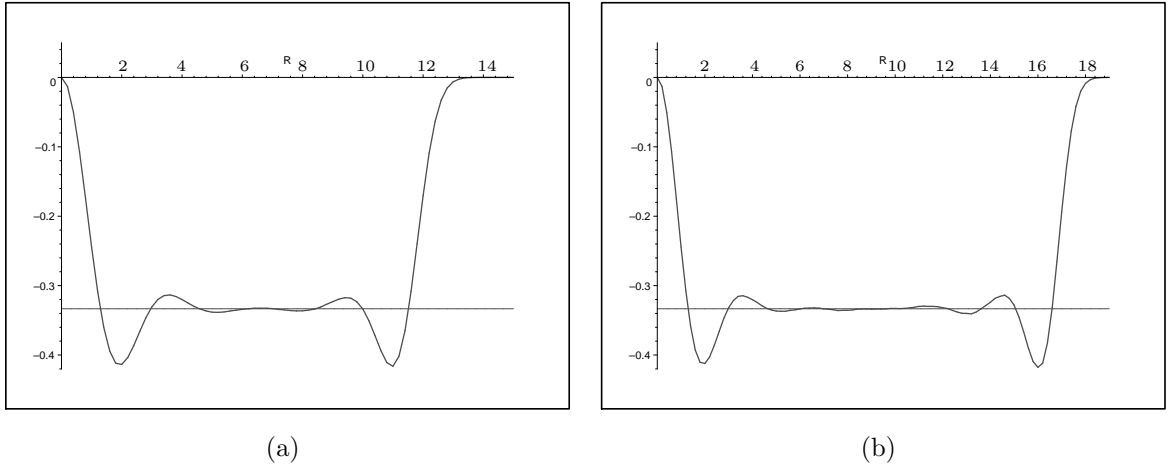

\begin{center}
\begin{tabular}{cc}
\subfigure[]{\psfig{figure=Lhlaoversikt1A.ps,angle=270,width=7.5cm}} &
\subfigure[]{\psfig{figure=Lhlaoversikt1B.ps,angle=270,width=7.5cm}}
\end{tabular}
\put(-402,67){$_{_2}$}
\put(-380,67){$_{_4}$}
\put(-356,67){$_{_6}$}
\put(-334,67){$_{_8}$}
\put(-312,67){$_{_{10}}$}
\put(-289,67){$_{_{12}}$}
\put(-266,67){$_{_{14}}$}
\put(-182,67){$_{_2}$}
\put(-163,67){$_{_4}$}
\put(-146,67){$_{_6}$}
\put(-128,67){$_{_8}$}
\put(-111,67){$_{_{10}}$}
\put(-93,67){$_{_{12}}$}
\put(-76,67){$_{_{14}}$}
\put(-58,67){$_{_{16}}$}
\put(-39,67){$_{_{18}}$}
\end{center}
\caption[]{\footnotesize Quasi-hole charge $C_{\rm h}(R)$
for systems with 50 and 100 electrons. The horizontal line is $-1/3$.}
\label{fig2}
\end{figure}

\begin{figure}[htb]
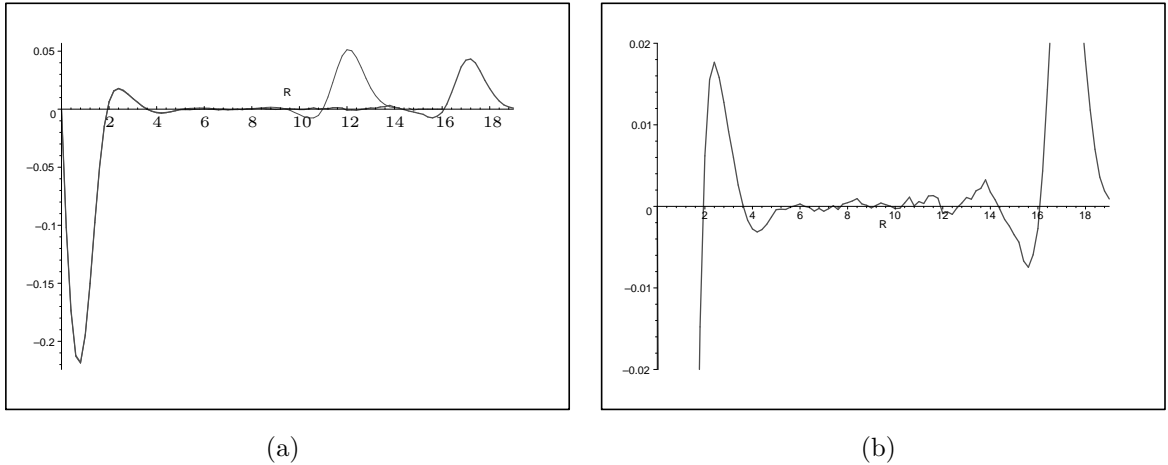

\begin{center}
\begin{tabular}{cc}
\subfigure[]{\psfig{figure=Lhfloversikt1.ps,angle=270,width=7.5cm}} &
\subfigure[]{\psfig{figure=Lhfldetalj1.ps,angle=270,width=7.5cm}}
\end{tabular}
\put(-407,46){$_{_2}$}
\put(-389,46){$_{_4}$}
\put(-371,46){$_{_6}$}
\put(-353,46){$_{_8}$}
\put(-336,46){$_{_{10}}$}
\put(-318,46){$_{_{12}}$}
\put(-301,46){$_{_{14}}$}
\put(-282,46){$_{_{16}}$}
\put(-264,46){$_{_{18}}$}
\end{center}
\caption[]{\footnotesize (a): Quasi-hole charge fluctuations $D_{\rm
h}(R)$
for systems with 50 and 100 electrons.
(b): Enlarged vertical axis emphasizing the bulk behaviour of $D_{\rm
h}(R)$ for 100 electrons.}
\label{fig3}
\end{figure}

Fig.~\ref{fig2} shows the $R$-dependence of the quasi-hole charge
$C_{\rm h}(R)$, for system sizes of 50 and 100 electrons. The figure
shows that there are three distinct regions. For small $R$ there is a
region where the charge builds up when $R$ increases, which we clearly
may identify with the location of the quasi-particle. There is an
intermediate region where the charge seems to stabilize at a constant
value, which is consistent with the expected bulk value $-1/3$ for the
quasi-hole charge. For larger $R$ there is a region where the charge
again decreases to zero, and this we identify with the edge region of
the droplet. The charge profile of the quasi-hole, for small $R$, is
essentially identical for $50$ and $100$ electrons, and that is also
the case for the charge profile at the edge. The main difference
between the two cases is the size of the intermediate region, the bulk
region of the electron droplet.

The distinction between the three regions for $m=3$ is similar to what is
seen in Fig.~\ref{fig1}a for $m=1$ . The main difference is the presence
of  oscillations for $m=3$, in the charge profile for small and large
$R$, which to some extent extends into the intermediate region. These
oscillations are of the same form as has previously been found in
numerical calculations of the charge density of Laughlin quasi-particles
\cite{halpmorf}.

In Fig.~\ref{fig3} the charge fluctuations $D_{\rm h}(R)$ of the
Laughlin quasi-hole are shown. Figure (a) compares the results obtained
for systems with 50 and 100 electrons, whereas figure (b) shows an
enlarged picture of the 100 electron case. The irregularities of the
curve seen here are presumably due to the statistical fluctuations in
the Monte-Carlo calculations and they give an indication of the size of
these statistical errors. The results shown in Fig.~\ref{fig3} confirms
the picture that the effect of the quasi-hole is restricted to a limited
region around the origin. In the bulk of the electron droplet the charge
fluctuations vanish within small statistical errors, whereas there
are substantial fluctuations in the charge inside the quasi-hole and at
the edge.  According to the discussion in section \ref{massedef}, this
is consistent with the assumption that the charge of the Laughlin
quasi-hole is a sharply defined quantum number.

The results presented so far are obtained for a discrete set of
$R$-values, $R=hk$ with $h=0.2$ and  $k=0,1,\cdots,K$. The maximum
number $K$ is chosen  such that $hK$ is larger than the radius
$\sqrt{mN}$ of the electron droplet.  Recall that lengths are measured
in units of $\sqrt 2 \ell_B$, with $\ell_B$ as the magnetic length.
The ground state data are for the case of 50 electrons obtained
from 45 million electron configurations, and for 100 electrons from
17 million configurations. All quasi-hole data are found from
10 million configurations. For each data set we have numerically
estimated the standard  deviation of the considered mean value. This
enables us to estimate the numerical errors in the differences $C_{\rm
h}(R) = Q_{\rm h}(R) - Q_{\rm 0}(R)$ and $D_{\rm h}(R) = \Delta Q_{\rm
h}(R) - \Delta Q_{\rm 0}(R)$ as well. Table \ref{tab0} shows some of the
expectation values along with the computed statistical errors in the
case of 50 electrons. The results are consistent with a
quasi-hole charge that has the bulk value $-1/3$ and vanishing charge
fluctuations. The estimated values for the standard deviation are
somewhat larger than the irregularities of the plotted curve shown in
Fig.~\ref{fig3}.

\begin{table}[htb]
\centering
\begin{tabular}{|c|c|c|} \hline\hline
  $R $ &  $C_{\rm h}(R) $ &  $D_{\rm h}(R)$ \\
\hline $4.0$  & $ -0.3219 \pm 0.0003 $ & $ -0.0028 \pm 0.002 $ \\
\hline $6.0$  & $ -0.3334 \pm 0.0004 $ & $ +0.0003 \pm 0.005 $ \\
\hline $8.0$  & $ -0.3371 \pm 0.0005 $ & $ +0.0000 \pm 0.008 $ \\
\hline $10.0$ & $ -0.3345 \pm 0.0005 $ & $ -0.0044 \pm 0.013 $ \\
\hline\hline\hline
\end{tabular}
\caption{
\protect \footnotesize Charge $C_{\rm h}$ and
charge fluctuations $D_{\rm h}$ for the Laughlin quasi-hole in the case
of 50 electrons. Both mean values and estimated errors are listed. The
quantities $Q_{\rm h}$ and $\Delta Q_{\rm h}$ were obtained from 10
million electron configurations, while 45 million configurations were
used to find $Q_0$ and $\Delta Q_0$.}
\label{tab0}
\end{table}

\vspace{0.5cm}

\begin{figure}[htb]
\begin{center}
\psfig{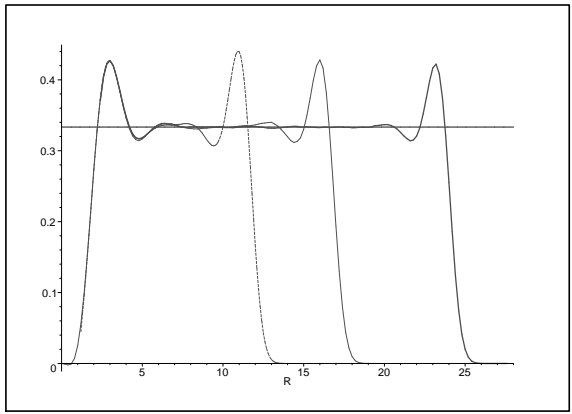}
\end{center}
\caption[]{\footnotesize Charge $C_{\rm Le}(R)$ of a Laughlin
quasi-electron. For $0 \leq R \leq 2.4$ the curve is obtained
by using Method 1 of section \ref{s33} for a system with 50 electrons.
For $R \geq 2.4$ Method 2 was used, and the curves are for 50,
100 and 200 electrons. The horizontal line is $1/3$.}
\label{fig7}
\end{figure}

We will now turn to the case of a Laughlin quasi-electron located at the
origin, and we consider first Fig.~\ref{fig7} where the charge $C_{\rm
Le}(R)$ is displayed. For $R\leq 2.4$ the curve is found for a system
with 50 electrons using  Method 1 of the previous section. The three
different curves for $R\geq 2.4$ are for 50, 100 and 200 electrons and
are found using Method 2. We observe again that there is a well defined
region where the value of the charge is almost constant and agrees with
the expected bulk value of $1/3$ for the integrated quasi-electron
charge. A more detailed presentation of the results show that for $100$
and $200$ electrons this is indeed the case within the small errors of
the calculation (to be discussed below) whereas for $50$ electrons
there is a small deviation which we assume to be a finite-size effect
(see Fig.~\ref{fig9}). Comparing with Fig.~\ref{fig2} we see that the
curves for $C_{\rm Le}(R)$ are, apart from a sign, similar to the curves
obtained for the quasi-hole.

\begin{figure}[htb]
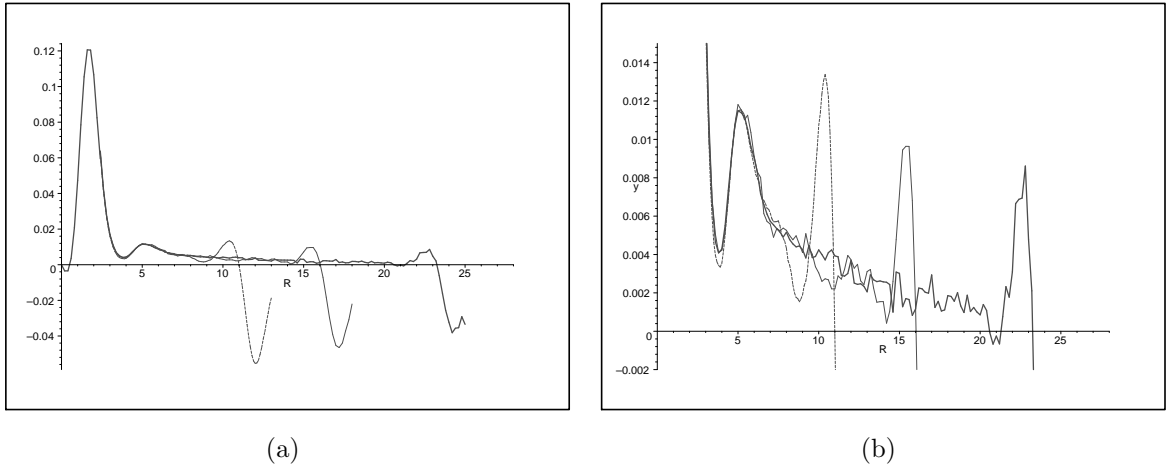

\begin{center}
\begin{tabular}{cc}
\subfigure[]{\psfig{figure=Lefloversikt1.ps,angle=270,width=7.5cm}} &
\subfigure[]{\psfig{figure=Lefldetalj1.ps,angle=270,width=7.5cm}}
\end{tabular}
\end{center}
\caption[]{\footnotesize Laughlin quasi-electron charge
fluctuations $D_{\rm Le}(R)$. The single curve for $0 \leq R \leq 2.4$
is found by using Method 1 of section \ref{s33} in a system with 50
electrons. The three curves for $R \geq 2.4$ are for 50, 100 and
200 electrons, and Method 2 was used.}
\label{fig8}
\end{figure}

Fig.~\ref{fig8} presents the calculated values of the charge
fluctuations $D_{\rm Le}(R)$ of the Laughlin quasi-electron. (Also here
the single curve shown for $R\leq 2.4$ is obtained for a system of 50
electrons with the use of Method 1, and for $R\geq
2.4$ the three curves are  obtained for systems of 50, 100 and 200
electrons with the use of Method 2.) Figure (b) is an
enlarged version of figure (a), now without the  small $R$
dependence.  We observe that for the system sizes considered here,
there are surviving fluctuations for the whole range of $R$-values
inside the electron droplet. This result is seen most clearly in
Fig.~\ref{fig8}b, where the irregularities indicate the size of
the statistical fluctuations in the computation. A further discussion
of uncertainties in the result will follow below.

\begin{figure}[htb]
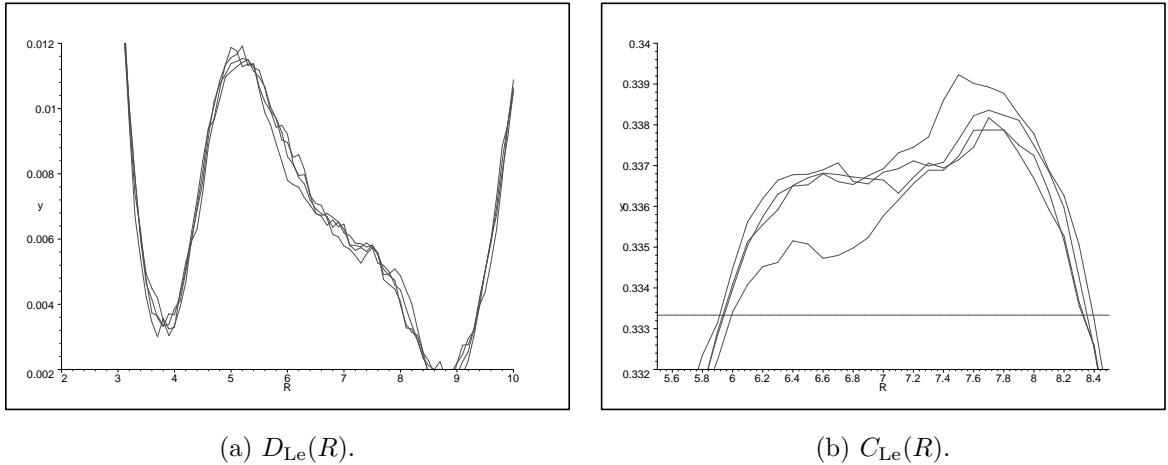

\begin{center}
\begin{tabular}{cc}
\subfigure[$D_{\rm
Le}(R)$.]{\psfig{figure=Leflerror1.ps,angle=270,width=7.5cm}} &
\subfigure[$C_{\rm
Le}(R)$.]{\psfig{figure=Lelaerror1.ps,angle=270,width=7.5cm}}
\end{tabular}
\end{center}
\caption[]{\footnotesize Four statistically independent curves for the
charge fluctuations $D_{\rm Le}(R)$, (a), and the charge $C_{\rm
Le}(R)$,
(b)
of a Laughlin
quasi-electron. Each curve is based on approximately $48$ million
configurations for the quasi-electron data. The same ground state
data,
found from 125 million configurations,
were used in all four cases. The lattice constant is $h=0.1$}
\label{fig9}
\end{figure}

The  behaviour of the charge fluctuation curve for the Laughlin
quasi-electron is clearly different from that of the quasi-hole, which
is shown in Fig.~\ref{fig3}. For a system of $100$ electrons
the quasi-hole charge fluctuations were seen to vanish in the bulk,
within the small statistical errors of the calculations.  A comparison
between Figs.~\ref{fig8}b and \ref{fig3}b, with compensation for the
difference in vertical scale, emphasizes the difference between the two
cases. We thus conclude that there are charge fluctuations for the
Laughlin quasi-electron that are larger and extend much further out
than they do for the quasi-hole. Even though the expectation value of
the quasi-electron charge is well defined and agrees with the expected
bulk value (within the statistical uncertainty) in an intermediate
interval of $R$, there are small but non-vanishing charge fluctuations
which persist also here. In this sense the charge is not a sharply
defined  quantum number for the quasi-electron. However, there is a
slow decay in the fluctuation curve for increasing $R$, and we cannot
rule out that for even larger systems, with electron numbers
$N>200$, the value of the charge fluctuation will settle at the value
zero further out in the bulk of the electron droplet. If that is the
case, the quasi-electron charge is a sharp quantum number, but the size
of the quasi-electron, as measured through these fluctuations, will be
much larger than the size estimated from the charge expectation value or
found by comparison with the charge fluctuations of the quasi-hole.

The curves shown here for $R\geq 2.4$  are based on the following
number of electron configurations: To find the quantities $Q_{\rm
Le}(R)$ and $\Delta Q_{\rm Le}(R)$ we used 191 million configurations
in the case of 50 electrons, 42 million for 100 electrons and 102
million for 200 electrons. The ground state  quantities $Q_0(R)$ and
$\Delta Q_0(R)$ were obtained from 125 million configurations for 50
electrons, 17 million for 100 electrons and 86 million electron
configurations in the case of 200 electrons. For $R\leq 2.4$ we used
4.4$\cdot$10$^9$ electron configurations to find the quasi-electron
quantities $Q_{\rm Le}(R)$ and $\Delta Q_{\rm Le}(R)$. The calculations
were done for values of $R=hk$, with $k=0,1,\cdots,K$, and $h=0.1$ in
the 50 electron case and $h=0.2$ for the two other system sizes.  We
would also like to mention that in the case of 50 electrons, for a
small range of $R$ around $R=2.4$, both the calculation methods,
Method 1 and Method 2, converged well and gave  coinciding results for
$C_{\rm Le}(R)$ and $D_{\rm Le}(R)$.

\begin{figure}[htb]
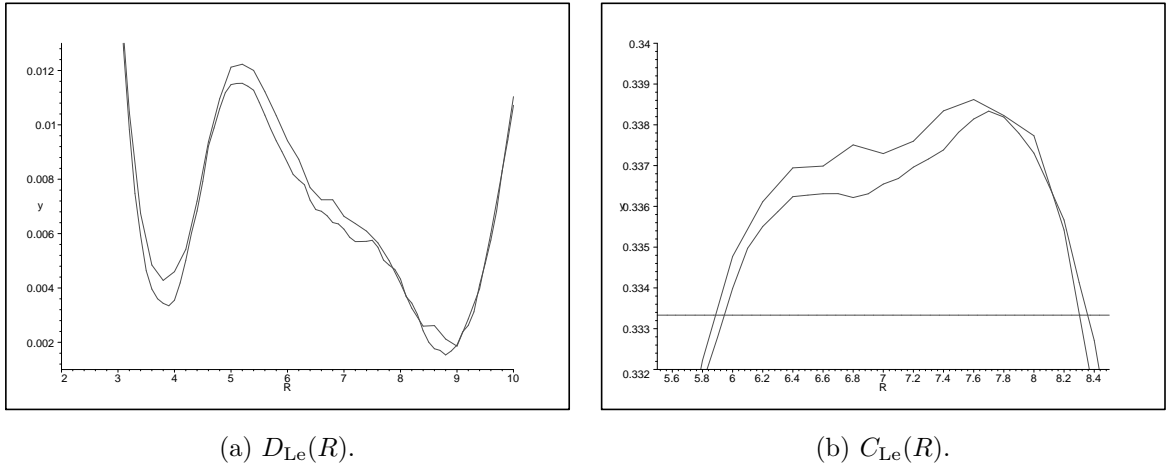

\begin{center}
\begin{tabular}{cc}
\subfigure[$D_{\rm
Le}(R)$.]{\psfig{figure=Leflerrorder1.ps,angle=270,width=7.5cm}} &
\subfigure[$C_{\rm
Le}(R)$.]{\psfig{figure=Lelaerrorder1.ps,angle=270,width=7.5cm}}
\end{tabular}
\end{center}
\caption[]{\footnotesize Charge fluctuations $D_{\rm Le}(R)$, (a), and
charge $C_{\rm Le}(R)$, (b), found from two different lattice spacings
$h=0.1$ and $h=0.2$ in the case of 50 electrons.}
\label{fig10}
\end{figure}

The conclusion for the charge fluctuations suggested above is based
on the assumption that the irregularities seen in the
fluctuation curves give a measure of the errors introduced in the
computation. However, in order to establish the conclusion more
firmly, we have considered numerical errors more carefully.
The errors include accidental errors due to statistical
fluctuations in the numerically obtained mean values, as well as
systematical errors due to the finite lattice spacing $h$ used in the
numerical derivatives. Since the charge fluctuations, as well as the
charge expectation values, have been calculated by a different
method, which involves ``negative probabilities", the standard
deviations of the calculated mean values are not so easily
established, as in the case of the quasi-hole. In the case of the
Laughlin quasi-electron the statistical errors therefore have been
estimated simply by comparing the results of independent runs of the
Monte-Carlo routine. To investigate the possible systematical errors
we compare the results obtained by the use of two different step
lengths and we compare the results of numerical differentiation with
exact results, by applying the methods to a representative test
function which can be handled analytically.

The estimate for the statistical errors is obtained by
dividing the numerical data for ${\widetilde Q}(R)$ and ${\widetilde
F}(R_1,R_2)$ into four statistically independent piles, each consisting
of approximately $48$ million electron configurations.
For each pile we have computed
$C_{\rm Le}(R)$ and $D_{\rm Le}(R)$, using the same ground state data
in all cases. Fig.~\ref{fig9}a shows the four fluctuation curves found
in this way for the case of $50$ electrons, while Fig.~\ref{fig9}b shows
the four independent charge  curves. For a given $R$ we estimate the
errors in the mean values as half the difference $(D_{\rm Le}(R))_{\rm
max} - (D_{\rm Le}(R))_{\rm min}$, and similarly for $C_{\rm Le}(R)$.
For the charge fluctuations this number is from the figure
seen to typically take the value $0.0005$, and implies
for example for $R=7$ that the mean value found is approximately
$10$ times larger than this error. Fig.~\ref{fig9} also shows that the
estimation of the errors obtained through the independent runs gives
essentially the same results as the estimation obtained visually from
the irregularities of a single curve. As judged from Fig.~\ref{fig8}
the statistical errors for larger $R$, in the $100$ and $200$ electron
curves, may be slightly larger than displayed by the $50$ electron
curves. However, the value of the fluctuations still are significantly
different from zero in most of the electron droplet.

A similarly estimated error in the charge expectation value $C_{\rm
Le}(R)$ is approximately $0.001$. A detailed presentation (not included
here) of the results in the cases of $100$ and $200$ electrons shows
that the calculated charge expectation values do not deviate
significantly (within this error) from $1/3$ in the bulk of the
electron droplet.

\begin{figure}[htb]
\begin{center}
\begin{tabular}{cc}
\subfigure[${\widetilde
F}(R_1,R_2)$]{\psfig{figure=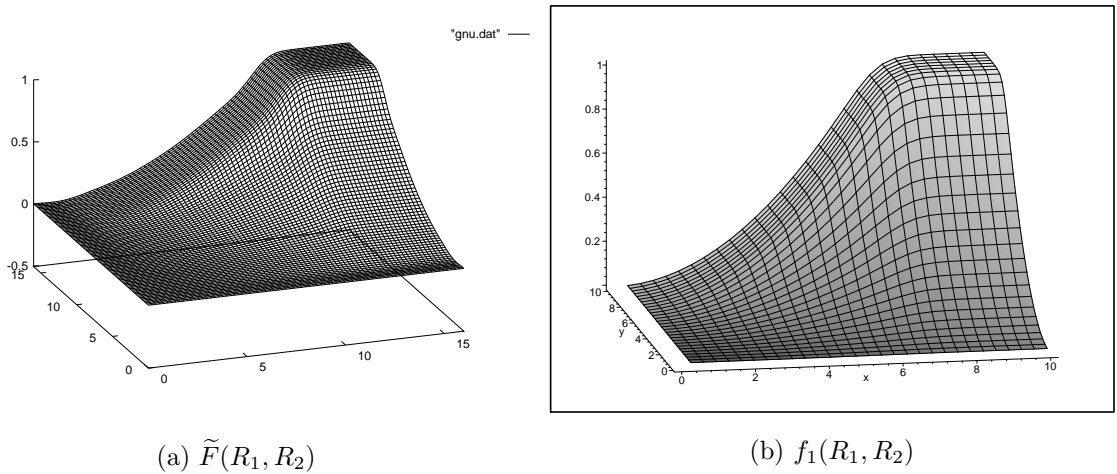,angle=270,width=7.5cm}} &
\subfigure[$f_{1}(R_1,R_2)$]{\psfig{figure=3Dtest.ps,angle=270,width=7.5cm}}
\end{tabular}
\end{center}
\caption[]{\footnotesize Comparison between (a): The Monte Carlo
estimated function ${\widetilde F}(R_1,R_2)$ for 50 electrons and lattice
spacing
$h=0.2$, and (b): The analytically found test function
$f_{1}(R_1,R_2)$ from
Eq.~(\ref{tf}), also calculated for 50 electrons. In both cases the
factor $N(N-1)$ has been divided out.}
\label{fig11}
\end{figure}

We will now discuss the systematical errors introduced by the finite
lattice spacing $h$. For small $h$ the lattice spacing will give
errors in the numerical derivatives, Eqs.~(\ref{tilq}) and
(\ref{tilf}), which are proportional to $h^2$. The above results for
50 electrons are all found using $h=0.1$. In addition we have
performed computations with $h=0.2$, and Fig.~\ref{fig10} compares
curves for $D_{\rm Le}(R)$ and $C_{\rm Le}(R)$, obtained with these
two values of $h$. We observe that the differences are similar in
size to the statistical error that we have already considered. Due
to the $h$-dependence of the error for small $h$ we expect that a
further reduction in the lattice spacing will introduce
differences which are smaller than the differences
between the results obtained with $h=0.2$ and $h=0.1$.
Results for $h=0.02$ seem to confirm this.

To make an independent check of the error introduced by the discrete
differentiation, we have applied this operation to a test function
with a similar form as ${\widetilde F}(R_1,R_2)$. From
Eqs.~(\ref{fakden}), (\ref{tilF}) and (\ref{laugr}) we observe that,
apart from the factor $\prod_{i=1}^N(|z_i|^2 -1)$ the function
${\widetilde F}(R_1,R_2)$ is nothing else than the integrated two-particle
density of the ground state for filling fraction $1/3$. We therefore
expect that except for a scaling in distance, the function
\beq
f_{1}(R_1,R_2) = \frac{N(N-1)}{I_{1}} \int_{A_1} {\rm d}^2 z_1
\int_{A_2} {\rm d}^2 z_2 \int {\rm d}^{2(N-2)} z \ |\psi_{1}|^2,
\label{tf}
\eeq
will mimic the main properties of the function ${\widetilde F}(R_1,R_2)$
reasonably well. Here $\psi_{1}$ is the wave function for the
$m=1$ ground state (\ref{laugr}), and $I_{1}$ is
the corresponding normalization integral. That this anticipation is
indeed fulfilled is seen in Fig.~\ref{fig11},
where the analytically determined function $f_{1}(R_1,R_2)$ is
compared to the numerically found function ${\widetilde F}(R_1,R_2)$.
Both cases are for 50 electrons, and in each figure the surface
starts from $0$ on the
$R_1$ and $R_2$ axes, and then smoothly builds up to the value
$N(N-1)$ as the edge of the electron droplet is reached.
For the test function $f_{1}(R_1,R_2)$ we have then
computed the right hand side of Eq.~(\ref{tilf}) in two ways:
We have computed it analytically, and in addition we have calculated
it by use of the discrete differentiation, Eq.~(\ref{dert}), with
lattice spacing $h=0.1$. We found errors in the result obtained with the
discrete derivatives which were approximately $ 0.0005$ in the
central part of the droplet. This agrees well with the results found
from runs of the Monte Carlo routine with the two different spacings,
$h=0.2$ and $h=0.1$, and are similar to the estimated value for the
statistical errors.

Based on these estimates of the errors introduced in the computation
we find it reasonable to conclude that the deviations from zero
seen in Fig.~\ref{fig8} for the charge fluctuations are real and not
due to the errors introduced by the calculations.

\begin{figure}[htb]
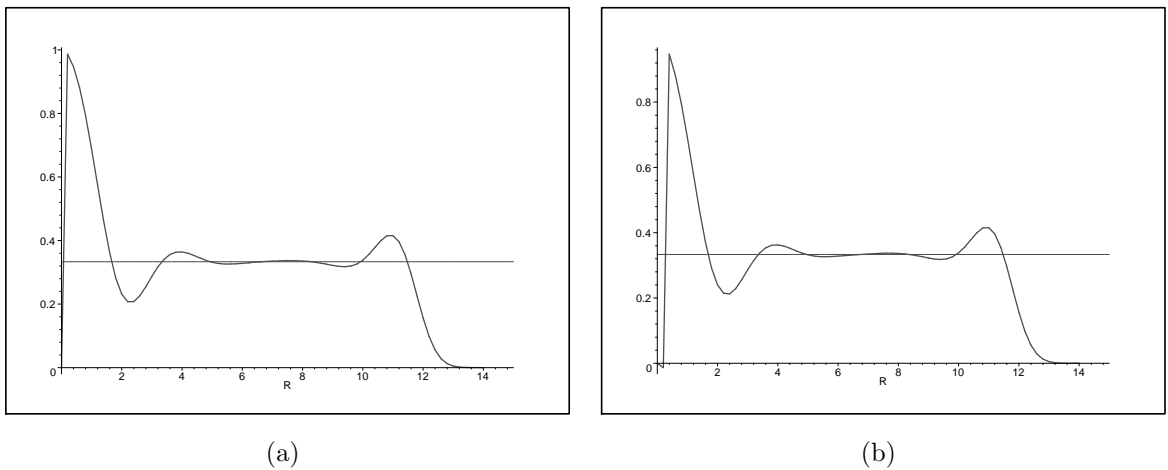

\begin{center}
\begin{tabular}{cc}
\subfigure[]{\psfig{figure=Jelaprojoversikt1.ps,angle=270,width=7.5cm}}
&
\subfigure[]{\psfig{figure=Jelauprojoversikt1.ps,angle=270,width=7.5cm}}
\end{tabular}
\end{center}
\caption[]{\footnotesize Charge $C_{\rm Je}(R)$ of a Jain
quasi-electron in a system with 50 electrons. (a): Projected state,
Eq.~(\ref{pjain}). (b):
Unprojected state, Eq.~(\ref{jainq}).}
\label{fig4}
\end{figure}

\begin{figure}[htb]
\begin{center}
\begin{tabular}{cc}
\subfigure[]{\psfig{figure=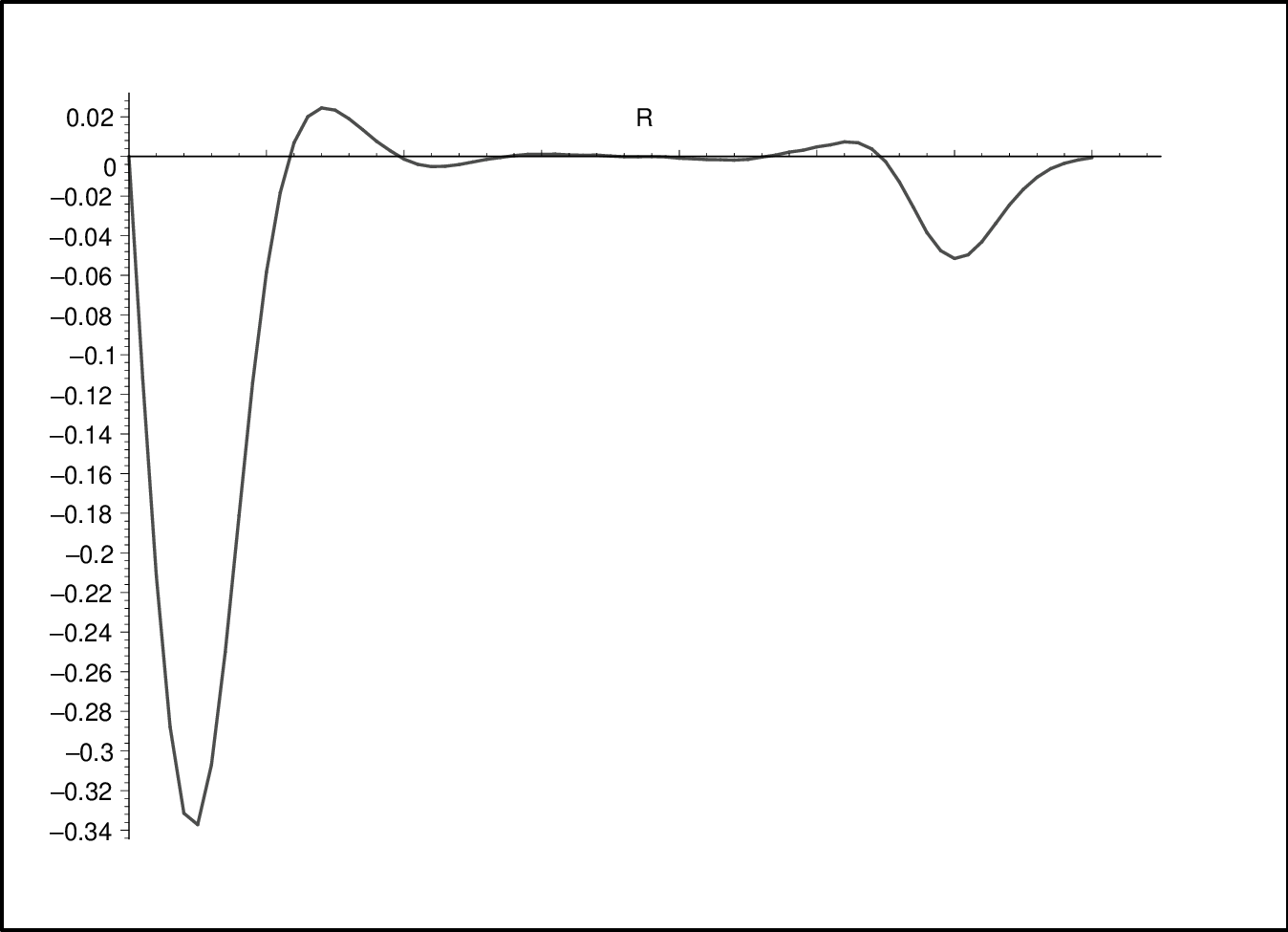,angle=270,width=7.5cm}}
&
\subfigure[]{\psfig{figure=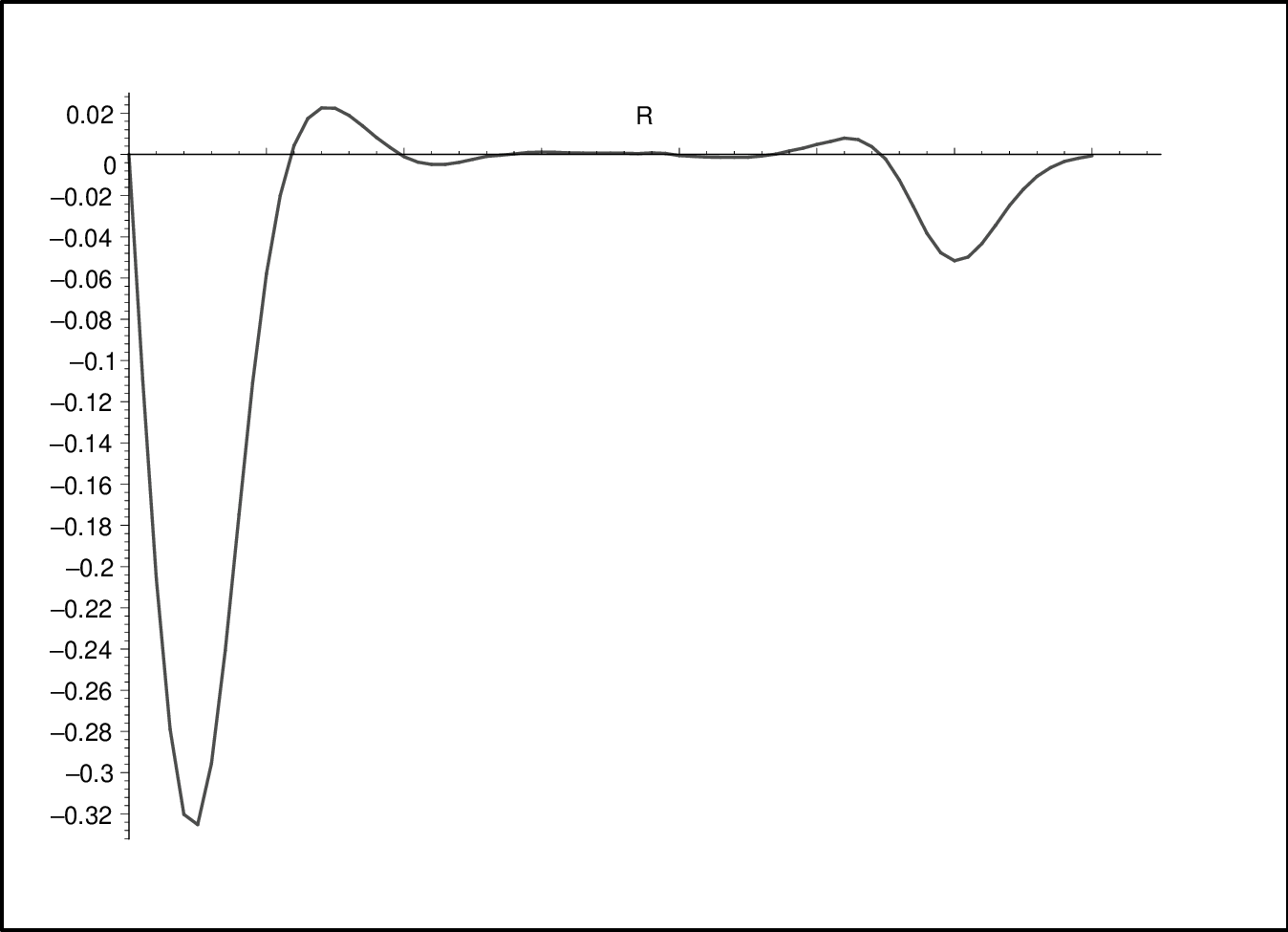,angle=270,width=7.5cm}}
\end{tabular}
\put(-404,70){$_{_2}$}
\put(-382,70){$_{_4}$}
\put(-356,70){$_{_6}$}
\put(-333,70){$_{_8}$}
\put(-312,70){$_{_{10}}$}
\put(-289,70){$_{_{12}}$}
\put(-266,70){$_{_{14}}$}
\put(-175,70){$_{_2}$}
\put(-153,70){$_{_4}$}
\put(-131,70){$_{_6}$}
\put(-109,70){$_{_8}$}
\put(-87,70){$_{_{10}}$}
\put(-65,70){$_{_{12}}$}
\put(-41,70){$_{_{14}}$}
\end{center}
\caption[]{\footnotesize Charge fluctuations $D_{\rm Je}(R)$ of a Jain
quasi-electron in a system with 50 electrons. (a): Projected state.
(b):
Unprojected state.}
\label{fig5}
\end{figure}

\begin{figure}[htb]
\begin{center}
\begin{tabular}{cc}
\subfigure[]{\psfig{figure=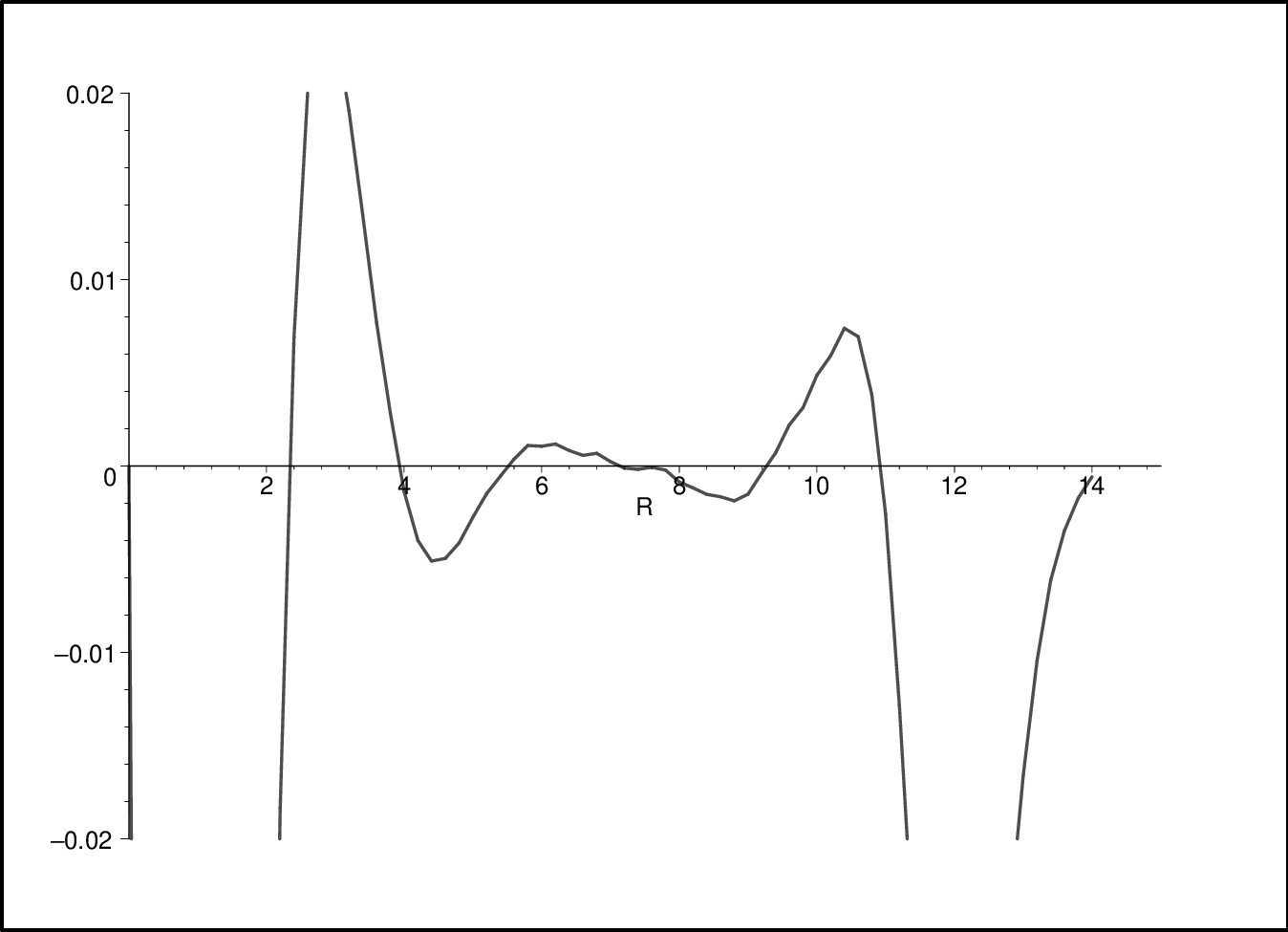,angle=270,width=7.5cm}}
&
\subfigure[]{\psfig{figure=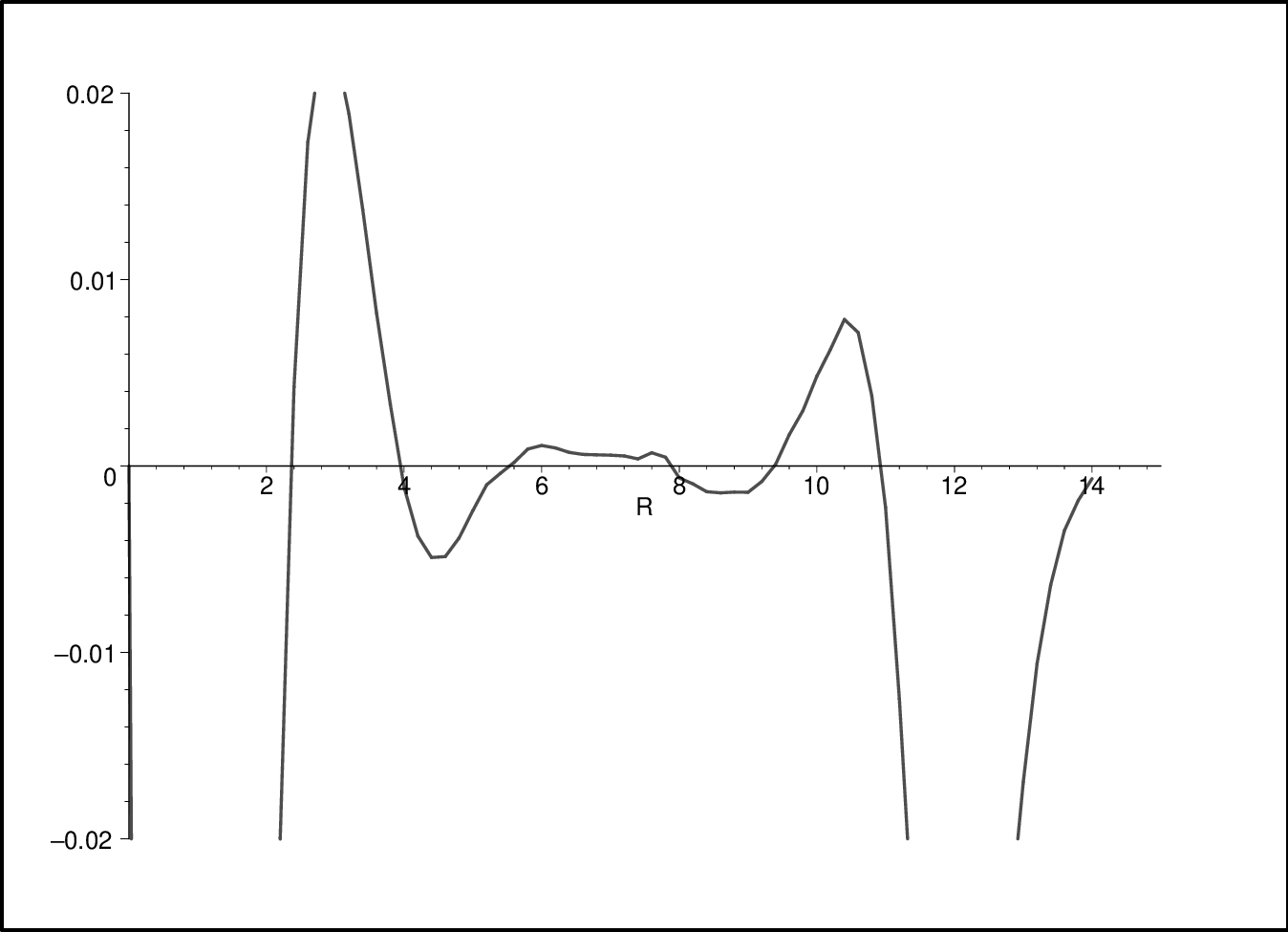,angle=270,width=7.5cm}}
\end{tabular}
\end{center}
\caption[]{\footnotesize Fluctuations $D_{\rm Je}(R)$ of a Jain
quasi-electron in a system with 50 electrons. (a): Projected state.
(b):
Unprojected state.}
\label{fig6}
\end{figure}

\vspace{0.5cm}

As noted above, we have also studied the charge and charge fluctuations
for the quasi-electron defined by Jain's wave function. This
investigation was prompted by the behaviour found for the Laughlin
quasi-electron, as a wish to compare the different proposals for the
wave function. For a Jain quasi-electron located at the origin
calculations have been performed for the projected state,
Eq.~(\ref{pjain}), as well as for the unprojected state,
Eq.~(\ref{jainq}). In both cases we have considered only a system of 50
electrons.

Fig.~\ref{fig4} compares the charge $C_{\rm Je}(R)$, as found for the
projected state, to the corresponding quantity for the unprojected
state, with the results for the projected one in the figure (a). We
readily notice that there is almost no difference between the two
figures,
\ie projection onto the lowest Landau level does not make any
important difference in the present context. In both cases there is a
well defined intermediate region where the curve is consistent with
the expected integrated charge value $1/3$ (times the electron
charge). When compared to Fig.~\ref{fig7} we notice that the amplitude
of the peak for small $R$ is larger for the Jain quasi-electrons than
it is for the Laughlin quasi-electrons, but apart from this the
differences in charge expectation values are rather small for the two
definitions of the quasi-electron wave function.

This similarity changes when we consider the charge fluctuations. For
the Jain quasi-electron these are displayed in  Fig.~\ref{fig5}, and
Fig.~\ref{fig6} shows the same results with the vertical axes
enlarged. We observe again that the projection onto the lowest Landau
level has a very small effect on the result. But the most compelling
observation is that there exists a small region, which we interpret as
corresponding to the bulk of the droplet, where the charge fluctuations
are very small, indeed consistent with the value zero within small
statistical errors (see the discussion below). The range of
$R$ for which this is true is about the same as the range where the
charge mean value, shown in Fig.~\ref{fig4}, is close to the bulk
value $1/3$. This behaviour is similar to the results we found for the
Laughlin quasi-hole, and differs from what we saw for the Laughlin
quasi-electron.

\begin{table}[htb]
\centering
\begin{tabular}{|c|c|c|} \hline\hline
  $R $ &  $C_{\rm Je}(R) $ &  $D_{\rm Je}(R)$ \\
\hline $4.0$  & $ 0.3640 \pm 0.0002 $ & $-0.001 \pm 0.001 $ \\
\hline $6.0$  & $ 0.3281 \pm 0.0002 $ & $+0.001 \pm 0.003 $ \\
\hline $8.0$  & $ 0.3355 \pm 0.0003 $ & $-0.001 \pm 0.005 $ \\
\hline $10.0$ & $ 0.3383 \pm 0.0003 $ & $+0.005 \pm 0.007 $ \\
\hline\hline\hline
\end{tabular}
\caption{
\protect \footnotesize Charge $C_{\rm Je}$ and
charge fluctuations $D_{\rm Je}$ for the Jain quasi-electron in
the case of 50 electrons.
Both mean values and estimated errors are listed.}
\label{tab1J}
\end{table}

The results presented here are based on the following set of data:
For the ground state we used 45 million electron configurations,
whereas we for the Jain quasi-electrons used 77 million
configurations for the projected state and 34 million for the
unprojected. We have performed calculations for $R=hk$ with $h=0.2$ and
$k=0,1\cdots,K$. For each set of data we estimated the numerical
standard deviations in $C_{\rm Je}$ and $D_{\rm Je}$. Some of the
results are listed in Table
\ref{tab1J}. The table shows, for the listed values of $R$, that the
calculated values for both $C_{\rm Je}$ and $D_{\rm Je}$ are consistent
with the expected bulk values within the estimated errors.

\section{Statistics parameter of the Jain quasi-electrons}

The studies of the charge fluctuations presented in this paper have been
motivated by the results of Ref.~\cite{Kjonsberg98}, where the charge
and statistics parameters of the Laughlin quasi-electrons were studied
numerically, by Berry phase calculations, for systems of up to $200$
electrons. Whereas the expected charge value, in those calculations, was
well reproduced, no stable value for the statistics parameter was found.
This was clearly different from  calculations of the statistics
parameter of the quasi-hole. Our hypothesis has been that the
discrepancy between the quasi-hole and quasi-electron results is
due to more long range fluctuations in the correlation functions for
the latter. This conclusion seems to be supported by the results of
the charge fluctuation calculations presented in this paper.

The alternative definition of the quasi-electron wave function, given
by Jain, seems to be better behaved as far as the charge fluctuations
are concerned. Although we have not examined this wave function in an
equally detailed way as the one introduced by Laughlin, this seems to  be
a reasonable conclusion, based on the results of the calculations on
the $50$ electron system. Thus, the fluctuations relative to those of
the ground state, when moving away from the center of the quasi-electron,
seem to be more rapidly damped for Jain's wave function than
for the wave function defined by  Laughlin. To pursue this point further
we will now present the results of numerical calculations of the charge
and statistics parameter, as extracted from Berry phases, in the case of
Jain's quasi-electron. The results presented are for a system of 74
electrons at filling fraction $1/3$, and for simplicity we have used the
unprojected quasi-electron wave function. We will show that in this
case a rather well-defined value is found both for the charge and
statistics parameters. The value of the charge parameter is close to the
expected value $1/3$, and the statistics  parameter is close to $-1/3$.
The sign of the latter is the opposite of that of the quasi-hole and it
is also the opposite of what is expected from a simple model of the
quasi-electron as a charge-flux composite.

Let us briefly review how charge and statistics parameters are extracted
from Berry phases. The idea, as originally set forth by Arovas et.al. in
Ref.~\cite{Arovas84}, is to let the position parameter $z_0$ that
labels the quasi-particle state traverse a loop in the plane. The Berry
phase \cite{berry} associated with this motion can then be computed.
This phase is in turn interpreted as an Aharonov-Bohm phase
\cite{Aharonov59} for the unknown charge $q$ of the quasi-particle
encircling the known magnetic flux, and the charge is extracted. Thus,
there is a non-trivial interpretation that lies under this way of
determining the charge. To find the statistics parameter one computes
the Berry phase associated with two quasi-particles encircling one
another, and interprets the two-particle contribution to the Berry
phase as an anyon interchange parameter.

To be more specific, let $|z_0\rangle$ be the normalized state
corresponding to a single quasi-particle located at the
position $z_0$, and suppose the particle is moved around a closed
loop parameterized by $z_0 = re^{i\phi}$ with $\phi$ running from $0$ to
$2\pi$. The Berry connection is then defined by
\beq
A_1(r) = i \langle z_0|
\partial_{\phi}|z_0\rangle,
\label{abst}
\eeq
and the Berry phase is the integral of the Berry
connection along the path. This phase we relate to the Aharonov-Bohm
phase for a charged particle in a (uniform) magnetic
field. If the path is left handed relative to the direction of the
magnetic field, as is the case in our calculations, the Aharonov-Bohm
phase associated with the propagation of the charge around the loop
 is
\beq
\gamma = 2\pi qr^2.
\eeq
Charge is here measured in units of the electron charge, and
$r$ is the dimensionless radius, measured in units of
$\sqrt{2}\ell_B$, of the circular loop.
The charge of the quasi-particle is now determined by setting
the Berry phase equal to the Aharonov-Bohm phase. If the Berry
connection depends on $r=|z_0|$ but not on $\phi$ (due to rotational
invariance), the Berry phase is
$\beta_{\rm 1}(r)=2\pi A_1(r)$
and the charge parameter is\beq
q(r) =  \frac{1}{r^2} \frac{\beta_{\rm 1}(r)}{2\pi r^2}. \label{lad}
\eeq
In general this definition will give an $r$-dependent charge  for
the quasi-particle, but far from the edge of the electron droplet $q(r)$
is expected to settle at a constant value.

Suppose there are two quasi-particles in the system, with
positions $\pm z_0$, so that the parameterization $z_0 = r
e^{i\phi}$,
with $\phi$ now running from $0$ to $\pi$,
describes a counterclockwise interchange of the two particles. We can
define the Berry connection associated with this interchange as
\beq
A_2 = i \langle z_0,-z_0|
\partial_{\phi}|z_0,-z_0\rangle,
\eeq
where $|z_0,-z_0\rangle$ is the two quasi-particle state. Subtracting
the single-particle contributions we have that $\nu$,
defined as
\beq
{\nu}(r) = \frac{1}{\pi}(\beta_{\rm 2}(r) -2 \beta_{\rm 1}(r)),
\label{anyon}
\eeq
with $\beta_2$ as the integral of $A_2$, can be identified (for large
separation $r$) with the anyon statistics parameter of the particles
\cite{Hansson92}.
Eqs.~(\ref{lad})
and (\ref{anyon}) constitute the basis for the discussion below.

Before proceeding, we would like to make a comment on the sign
convention used in the definition of the statistics parameter
(\ref{lad}), since the sign is of some importance for the discussion. The
sign is fixed by orienting the loop used for the Berry phase calculations
in a positive direction relative to the product $eB$ of the electron
charge and the external magnetic field. Thus, it is independent of the
sign of the charge $qe$ of the quasi-particle.  However, another
convention is possible, and maybe even more useful. If the orientation
of the loop in the calculation of the statistics parameter is fixed
relative to $qeB$ rather than $eB$, then the sign of the statistics
parameter has a direct physical significance. We may write the parameter
with this new convention as
\beq
\nu_1=-\frac{\beta_1}{\left| {\beta _1} \right|} \nu
\eeq
The sign of $\nu_1$ is then determined by the relative sign between
one-particle and the two-particle contributions to $\beta_2$. If
$\nu_1$ is positive there is an effective repulsive (statistical)
interaction between the two quasi-particles, and if it is negative there
is an attractive interaction\footnote {For anyons in the lowest Landau
level the parameter is normally restricted to the interval $0\leq \nu_1
<2$. However, there is a natural extension of this to the interval $-1<
\nu_1 <\infty$. Negative values then corresponds to anyons with an
additional attraction which gives a singular (but normalizable) short
range behaviour of the wave function. If $\nu_1$ is larger than $1$
there is a repulsion corresponding to the exclusion of one or more of
the lowest relative angular momentum states. (The Laughlin states of
filling fraction $1/m$ are of this kind.) The statistics parameter
$\nu_1$ with the given sign convention is identical (up to a constant
shift) to the one-dimensional statistics parameter introduced in terms
of algebraic relations between observables of the system \cite{Hansson92}. 
It is also identical to the exclusion statistics parameter defined by state
counting in the many quasi-particle space \cite{Haldane}.}.

A Jain quasi-electron located at the origin is described by the
wave function of Eq.~(\ref{detwave}). To find the charge and statistics
parameters we need to translate the quasi-electron to the position $z_0$
without moving the circular electron droplet itself. Since the
expression for such a translated quasi-electron (to the best of our
knowledge) does not exist in the literature, and since it does not
follow trivially from (\ref{detwave}), we will include a discussion of
how to solve this problem.

Notice that apart from the projection operator ${\cal P}$, the wave
function (\ref{detwave}) for $m=1$ describes a filled lowest Landau
level with a single electron pushed up to the next Landau level, where
it occupies the single-electron coherent state localized at the origin
\cite{Kjonsberg96} . If this single-electron state is translated to the
point $z_0$ it is described by
\beq
f_{z_0}(z,z^*) = \frac{1}{{\sqrt \pi}}(z^*-z_0^*)e^{zz_0^*}
e^{-\half (zz^* + z_0 z_0^*)}. \label{trcoh}
\eeq
Here $z$ is the electron coordinate, and the wave function
is normalized to unity in the state space of a single electron.
The coherent states are
known to be maximally localized, so if one of the $N$ considered
electrons now occupy $f_{z_0}(z,z^*)$
the
$m=1$ Jain wave function will
have an excess charge equal to the electron charge accumulated close
to the
position $z_0$. Hence the quasi-electron has been moved from the
origin to $z_0$.

\begin{figure}[htb]
\begin{center}
\psfig{figure=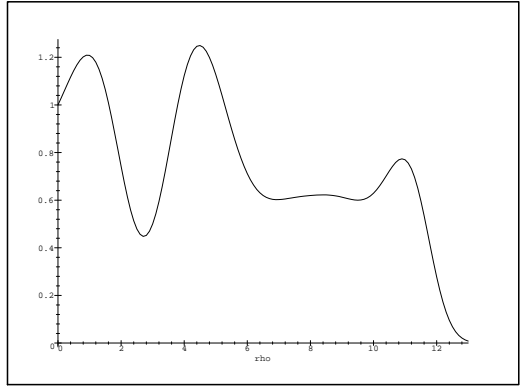,angle=270,width=7.5cm}
\end{center}
\caption[]{\footnotesize Cut in the radial direction of the single
electron density $\rho(z_1,z_1^*)$,
Eq.~(\ref{singe}),
for a Jain quasi-electron. The cut starts at the origin and goes
through the parameter point $z_0=3$. System size is 50 electrons. For
computational simplicity the density is not normalized.}
\label{fig12}
\end{figure}

For $m\neq 1$ the situation is a bit more complicated. The VanderMonde
determinant
\beq
\Delta = \prod_{k<l}(z_k - z_l)
\eeq
raised to the power $(m-1)$, which is the new factor in the wave
function as compared to the $m=1$ function, will push the electrons
apart. The detailed effect of this pushing is not completely known, but
we notice that $\Delta$ treats all electrons symmetrically. This implies
that a quasi-electron that for $m=1$ was located at $z_0$ now is moved,
assuming that we keep the form in Eq.~(\ref{trcoh}) of the coherent
state.
We do not have a simple argument to determine how the quasi-electron
position will depend on $m$, but numerical studies show that the new
position is close to $m z_0$.
This scaling with $m$ of the quasi-electron position can be avoided if
we replace
$z_0$ with $z_0/m$ in Eq.~(\ref{trcoh}). Hence for a Jain
quasi-electron
localized at the position $z_0$ we use the wave function
\be
\psi_{\rm Je}^{\rm 1} &=& e^{-\half\sum_{i=1}^N|z_i|^2}
\Delta^{m-1}\left| \begin{array}{cccc}

(z_1^*-\frac{z_0^*}{m})e^{\frac{z_0^*}{m} z_1} &
(z_2^*-\frac{z_0^*}{m})e^{\frac{z_0^*}{m} z_2} &\cdots &
(z_N^*-\frac{z_0^*}{m})e^{\frac{z_0^*}{m} z_N} \\
                                     1  & 1  &  \cdots &1     \\
                                    z_1 & z_2 & \cdots & z_N  \\
                                   \vdots & \vdots & \vdots & \vdots
\\
                                  z_1^{N-2} & z_2^{N-2} & \cdots &
z_N^{N-2}
                                          \end{array} \right|
\label{jatr}
\ee
That this wave function indeed has excess charge localized around the
position $z_0$ is shown in Fig.~\ref{fig12}. The figure shows a cut
from the origin of the electron droplet and
through the point $z_0=3$, of the
(non-normalized) single electron density\footnote{The numerical
method used to find $\rho(z_1,z_1^*)$ will be describe below.}
\beq
\rho(z_1,z_1^*) = C \int {\rm d}^{2(N-1)}\! z \,
|\psi_{\rm Je}^{\rm 1}(z_1,\cdots, z_N)|^2 ,  \label{singe}
\eeq
with $C$ as a constant. We see that the
density profile, with its dip close
to
$z_0$ and two peaks on each side of the dip mimics that of a
coherent state in the first Landau level, Eq.~(\ref{trcoh}), localized
at
$z_0=3$. This figure supports the definition we have used for the wave
function describing a translated quasi-electron.

For two quasi-electrons at the positions $\pm z_0$ we use
the wave function
\be
\psi_{\rm Je}^{\rm 2} &=& e^{-\half\sum_{i=1}^N|z_i|^2}
\Delta^{m-1}\left| \begin{array}{cccc}

(z_1^*-\frac{z_0^*}{m})e^{\frac{z_0^*}{m} z_1} &
(z_2^*-\frac{z_0^*}{m})e^{\frac{z_0^*}{m} z_2} &\cdots &
(z_N^*-\frac{z_0^*}{m})e^{\frac{z_0^*}{m} z_N} \\

(z_1^*+\frac{z_0^*}{m})e^{-\frac{z_0^*}{m} z_1} &
(z_2^*+\frac{z_0^*}{m})e^{-\frac{z_0^*}{m} z_2} &\cdots &
(z_N^*+\frac{z_0^*}{m})e^{-\frac{z_0^*}{m} z_N} \\
                                     1  & 1  &  \cdots &1     \\
                                    z_1 & z_2 & \cdots & z_N  \\
                                   \vdots & \vdots & \vdots & \vdots
\\
                                  z_1^{N-3} & z_2^{N-3} & \cdots &
z_N^{N-3}
                                          \end{array} \right|.
\label{2jatr}
\ee
This function treats the two quasi-electrons symmetrically.

Before we proceed to the numerical results we need to establish the
relations between the desired Berry connections, given
by Eqs.~(\ref{lad}) and (\ref{anyon}), and the normalization
integrals of the two wave functions above. The state $|z_0\rangle$
in Eq.~(\ref{abst}) generally can be expressed as
\beq
|z_0\rangle = \frac{1}{{\sqrt I}} \sum_{k=0}^K (z_0^*)^k a_k
|k\rangle,
\label{expab}
\eeq
where $|k\rangle$ are orthonormal basis states, $a_k$ are expansion
coefficients and
\beq
I = \sum_{k=0}^K r^{2k} |a_k|^2;  \;\;\;\;\;r=|z_0|,
\eeq
is the normalization factor. The Berry connection can be related to $I$
by the expression
\beq
A(r) = r^2 \frac{\rm d}{{\rm d}r^2}\ln I.
\label{berno}
\eeq
This relation relies on the property that $I$ depends only
on the absolute value $r$ of $z_0$ and not its phase.
Since an expansion of the wave function $\psi_{\rm Je}^{\rm 1}$
as a power series in $z_0^*$ can be shown to be an expansion in
terms of orthogonal total angular momentum eigenstates, in exactly the
same way as in Eq.~(\ref{expab}), the relation (\ref{berno})
holds,
with
\beq
I_{\rm Je}^{\rm 1} = \int {\rm d}^{2N}\! z \,
|\psi_{\rm Je}^{\rm 1}(z_1,\cdots, z_N^*)|^2 .  \label{not1}
\eeq
For the two quasi-electron state
an expansion in $z_0^*$ is again an expansion
in terms of orthogonal
angular momentum eigenstates. However, the expansion now has the form
\beq
|z_0,-z_0\rangle = \frac{1}{\sqrt{I}} \sum_{k=0}^K (z_0^*)^{2k+1} a_k
|k\rangle.
\eeq
The lowest power of $z_0^*$ is not $0$ in this expansion, which means
that the
wave function $\psi_{\rm Je}^{\rm 2}$ contains an unphysical
singularity
for $z_0 =0$. This singularity should be removed, and we can
achieve this by using a complex
normalization factor. This changes the relation between the Berry
connection and the normalization integral. We find
\beq
A_2(r) = r^2 \frac{\rm d}{{\rm
d}r^2}\ln I_{\rm Je}^{2} -1,
\eeq
where
\beq
I_{\rm Je}^{\rm 2} = \int {\rm d}^{2N}\! z \,
|\psi_{\rm Je}^{\rm 2}(z_1,\cdots, z_N^*)|^2   \label{not2}
\eeq
is the usual normalization integral.
To summarize, the charge and statistics parameter are related to the
normalization integrals in Eqs.~(\ref{not1}) and (\ref{not2}) by
\be
q &=& \frac{\rm d}{{\rm d}r^2} \ln I_{\rm Je}^{\rm 1}
\label{chnor}\\
\nu &=& r^2 \frac{\rm d}{{\rm d}r^2} (\ln I_{\rm Je}^{\rm 2}
- 2 \ln I_{\rm Je}^{\rm 1})-1. \label{nunor}
\ee

We are now ready to present the numerical method used to find the
normalization integrals $I_{\rm Je}^{\rm 1}$ and $I_{\rm Je}^{\rm
2}$ as a function of $r\!=\!|z_0|$. Monte Carlo integration
is used, and for the generating probability density
$p(z_1,\cdots,z_N^*)$ used in the Metropolis algorithm we have the
following requirements: $p$ should represent the
integrands properly so that the numerical uncertainty becomes as
small as possible, and it should not depend on $z_0$. To satisfy these
criteria, we have used the
squared ground state wave function as the probability density, that is
\beq
p(z_1,\cdots,z_N^*) = \frac{1}{\sqrt I_0} e^{-\sum_{i=1}^N|z_i|^2}
|\Delta|^{2m}.
\eeq
Here $I_0 = \int {\rm d}^{2N} z e^{-\sum_{i=1}^N|z_i|^2} |\Delta|^{2m}$.
The wave function for a single quasi-electron, \ie the function
that enters the expression for
$I_{\rm Je}^{\rm 1}$, can be written as
\beq
\psi_{\rm Je}^{\rm 1} =
e^{-\half\sum_{i=1}^N|z_i|^2} \Delta^m \sum_{i=1}^N (z_i^* -
\frac{z_0^*}{m})  e^{\frac{z_0^*}{m} z_i}
(-1)^{i+1}\frac{M_i}{\Delta}, \label{exj}
\eeq
where $M_i$ is the determinant arising when we remove the first row
and the $i$'th column of the original determinant in
Eq.~(\ref{jatr}). The wave function for two
quasi-electrons can be similarly expanded, although now a
double sum will appear for the sum in Eq.~(\ref{exj}). The great
numerical
advantage we have achieved by our choice of probability density $p$
is that for each specific electron configuration, the number of
multiplications needed to find the ratio
$\frac{M_i}{\Delta}$ is proportional to $N$ rather than $N^2$, the
latter being typical for $M_i$ and $\Delta$ separately. Here $N$ is
the number of electrons. The reduction by the factor $N$ dramatically
reduces the required computer time.
We generate electron configurations
according to the probability density $p$ above. For $z_0=r=hk$,
with $h$ fixed and $k=0,1,\cdots,K$, we estimate $I_{\rm Je}^{\rm 1}(r)$
and $I_{\rm Je}^{\rm 2}(r)$. Notice that the two normalization
integrals are found from the same set of electron configurations, and
hence have correlated uncertainties. As
noticed in Ref.~\cite{Kjonsberg98} this reduces the numerical 
error when we
find $\nu$ from the difference in Eq.~(\ref{nunor}) since the numerical
errors have a tendency to cancel.

The numerical method used to compute the single electron density in
Eq.~(\ref{singe}) is basically the same. However, in this case the
coordinate $z_1$ as well as $z_0$ needs to be treated as a
parameter in the numerical calculations. In this case we have
therefore generated electron configurations by using
the square of the $(N-1)$-electron ground state wave function as the
probability density.

\begin{figure}[htb]
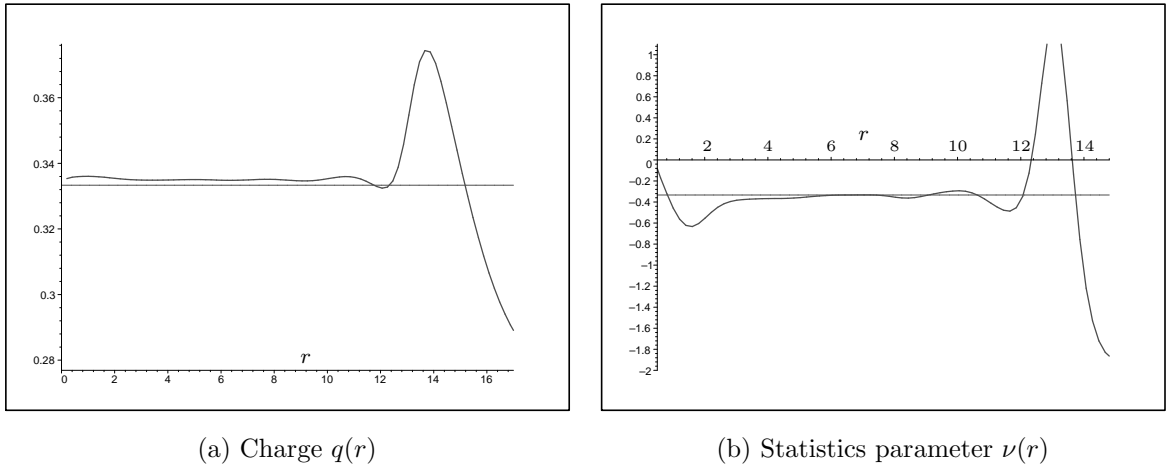

\begin{center}
\begin{tabular}{cc}
\subfigure[Charge
$q(r)$]{\psfig{figure=74eladn01.ps,angle=270,width=7.5cm}} &
\subfigure[Statistics parameter
${\nu}(r)$]{\psfig{figure=74estat01.ps,angle=270,width=7.5cm}}
\end{tabular}
\put(-181,37){$_{_2}$}
\put(-158,37){$_{_4}$}
\put(-134,37){$_{_6}$}
\put(-110,37){$_{_8}$}
\put(-88,37){$_{_{10}}$}
\put(-64,37){$_{_{12}}$}
\put(-40,37){$_{_{14}}$}
\put(-122,40){${_r}$}
\put(-333,-44){${_r}$}
\end{center}
\caption[]{\footnotesize (a): The charge $q(r)$ as a function of the
quasi-electron's distance from the origin. (b): The statistics parameter
${\nu}(r)$ as a function of half the distance between the
quasi-electrons. The system has 74 electrons, and for comparison we have
included the horizontal lines $1/3$ and
$-1/3$.}
\label{fig13}
\end{figure}

The results of our Berry phase calculations will now be presented. For
74 electrons we have determined the charge $q$ and statistics
parameter $\nu$ according to Eqs.~(\ref{chnor}) and
(\ref{nunor}). The computations are done for $m\!=\!3$, \ie for $1/3$
of a filled Landau level, and the required
integrals $I_{\rm Je}^{\rm 1}(r)$ and $I_{\rm Je}^{\rm 2}(r)$
are computed for the parameter $r$ taking the values $r = hk$, with
$h=0.2$ and $k=0,1,\cdots,K$.
Fig.~\ref{fig13} displays the results.

Fig.~\ref{fig13}a shows the charge
$q$ as given by Eq.~(\ref{chnor}) as a function of $r$, the
dimensionless distance from the
origin to the quasi-electron. We see that $q$ is (almost) 
constant and stays
close to the plateau
value $1/3$ all the way from the origin and until close to 
the edge of the
electron droplet.
For 74 electrons the latter has a radius of
$14.9$. This constant behaviour means that the bulk value of
the charge, as extracted from Berry phases, is well defined.
However, the small deviation from the expected value $1/3$ is
found to be statistically significant. It is interesting to
notice that a similar deviation was seen for the charge of the Laughlin
quasi-electron in Ref.~\cite{Kjonsberg98}. In that case it was
interpreted as a finite size effect that would vanish when
$N\rightarrow\infty$, since the deviation was seen to become smaller as
$N$ was raised. For the present case, a calculation for 50 electrons
gave the same value of the plateau as does Fig.~\ref{fig13}a, hence we
have no indication that the deviation is due to the finite size of the
electron droplet. We are aware that it may be related to the specific
way we have defined the wave function for the translated quasi-electron
in Eq.~(\ref{jatr}), but we have not studied this point thoroughly.

Notice that even for $r\rightarrow 0$, $q$ stays close to the
plateau rather than dropping to zero as does the integrated 
single-electron
density; compare to Fig.~\ref{fig4}. This emphasizes that the present
definition of the charge does not necessarily give the same charge as the
one
found by integrating
the single-electron density. When we interpret the Berry
phase as an Aharonov-Bohm phase, the behaviour as $r\rightarrow 0$ can
in the present case be understood in the
following way:
The Aharonov-Bohm phase $\gamma$ associated with the loop $z_0=r
e^{i\phi}$ is
proportional to
the encircled magnetic flux. The curve in Fig.~\ref{fig13}a shows
$\frac{1}{2\pi r^2}\gamma$, which then is a constant. This holds
even
if the charge is not truly point like, because all parts of the
smeared out charge is moved around in equal loops.

\begin{figure}[htb]
\begin{center}
\begin{tabular}{cc}
\subfigure[Statistics parameter for Jain
quasi-electrons.]{\psfig{figure=74estat01.ps,angle=270,width=7.5cm}} &
\subfigure[Statistics parameter for Laughlin
quasi-electrons.]{\psfig{figure=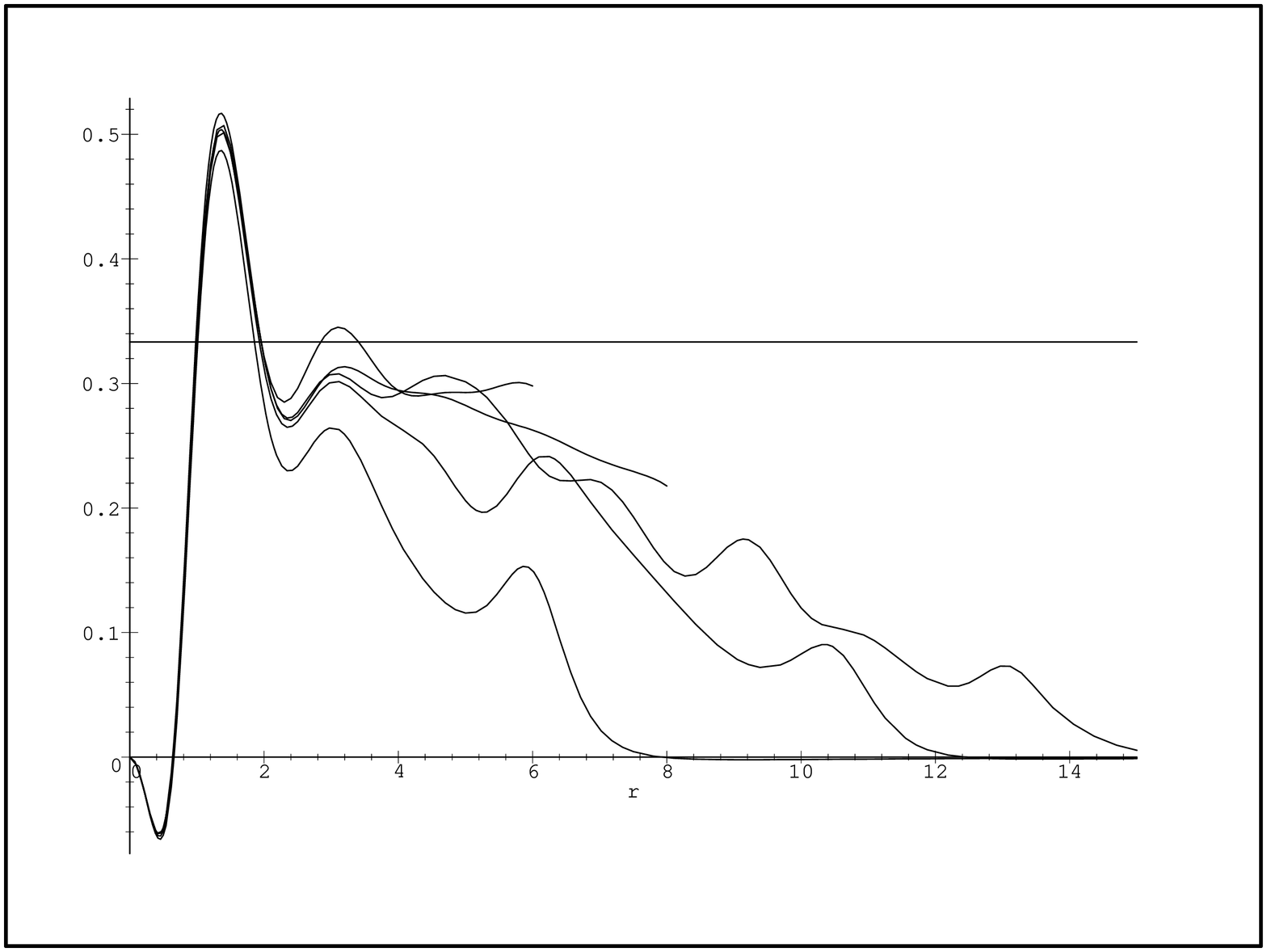,angle=270,width=7.5cm}}
\end{tabular}
\put(-407,42){$_{_2}$}
\put(-384,42){$_{_4}$}
\put(-360,42){$_{_6}$}
\put(-335,42){$_{_8}$}
\put(-313,42){$_{_{10}}$}
\put(-289,42){$_{_{12}}$}
\put(-266,42){$_{_{14}}$}
\put(-348,45){${_r}$}
\end{center}
\caption[]{\footnotesize Comparison between curves that define the
statistics parameters of (a): Jain quasi-electrons (\ref{nunor}), and
(b): Laughlin quasi-electrons. For
the Jain quasi-electrons we have used a system with 74 electrons,
whereas the Laughlin quasi-electron curves are for systems with 20, 50,
75, 100 and 200 electrons. Of these the 75 electron curve is farthest to
the right, and the figure is from Ref.~\cite{Kjonsberg98}. }
\label{fig14}
\end{figure}

We now turn to Fig.~\ref{fig13}b where the statistics parameter
${\nu}(r)$, as given by Eq.~(\ref{nunor}), is shown as a function of $r$.
The figure shows that there is a range of $r$ for which the value of
$\nu$ is approximately constant. That is, we see rather clear 
signs of a well
defined bulk value for the statistics parameter of the Jain
quasi-electrons. To give unquestionable evidence for the existence of
such a plateau, and also to
determine the precise value of the parameter, would require more
detailed studies than we have done here. However, we do observe that
the value $-1/3$ seems likely to be the value of the statistics
parameter $\nu$. This equals minus the statistics parameter of the
Laughlin quasi-holes. Let us mention that we also for a system with 50
electrons have seen clear indications of this.

The results seen here for the Jain quasi-electrons are similar to the
results obtained in Ref.~\cite{Kjonsberg98} for the Laughlin quasi-hole.
But the most striking observation is the big difference between the
results found in this reference for the Laughlin quasi-electron, and the
present results for the Jain quasi-electron. Fig.~\ref{fig14} shows a
comparison between the curves that are used to define the statistics
parameter, as found from Berry phases, for these two definitions of the
quasi-electron wave function, with Jain's definition used in figure (a)
and Laughlin's definition used in (b).
The latter figure is from Ref.~\cite{Kjonsberg98}, and shows
results for the system sizes 20, 50, 75, 100 and 200
electrons. The 75 electron curve, which is most appropriate for
comparison with our Jain quasi-electron
curve for 74 electrons, is the one
farthest to the right. The difference between Figs.\ref{fig14}a and
\ref{fig14}b is striking; whereas the curves for
Laughlin quasi-electron do not show signs of  a well-defined statistics
parameter, that is clearly the case for the Jain quasi-electron.

The signs of the statistics parameters of the quasi-electrons deserve
special attention. We note that they are different for Jain's and
Laughlin's wave functions. Let us re-express the results in terms of
the parameter $\nu_1$ which measures the repulsion or attraction between
the quasi-particles. For the quasi-hole the value is $\nu_1=+1/3$, which
corresponds to an effective repulsion between two quasi-holes. Compared
to the quasi-hole the Jain quasi-electron has the opposite sign for
the charge as well as for the statistics parameter $\nu$
\cite{Kjonsberg98}. This means that $\nu_1$ is positive also for the
Jain quasi-electrons. The opposite sign for the statistics parameter, as
may be indicated by the Berry phase calculations for the Laughlin
quasi-electron, would correspond to an effective attraction
between the quasi-electrons. In the anyon representation this attraction
would be represented by a singular anyon wave function
\cite{Kjonsberg98}.

A simple picture of the quasi-particles represented as
charge-flux composites indicates that (the non-integer part of) the
statistics parameter $\nu_1$ should have opposite sign for the
quasi-hole and the quasi-electron. This is because the sign of both the
flux and the charge is reversed for the quasi-electron compared to the
quasi-hole. This gives the same value for the parameter $\nu$, but the
opposite sign for $\nu_1$, due to the different signs of the charges.
This simple argument is supported by numerical studies of the
quantum Hall effect on a sphere \cite{Canright94}. State counting gives an
exclusion statistics parameter for the quasi-electrons which corresponds
to $\nu_1=2-1/3$, as compared to $\nu_1=+1/3$ for the quasi-holes. The
fractional part is negative, but it is compensated by a positive integer
part to give altogether a repulsive parameter. It is of interest to
note that the results obtained for Berry phase calculations of the
statistics parameter do not fit this expected value of the physical
quasi-electron, neither with Jain's nor Laughlin's wave functions for
the quasi-electron.

\section{Conclusions}

To summarize, we have examined expectation values and fluctuations
for the charge of quasi-holes and quasi-electrons in the quantum
Hall system with filling fraction $1/3$. The study has been motivated
by the asymmetry between quasi-holes and quasi-electrons that has
previously been found in Berry phase calculations of the statistics
parameter \cite{Kjonsberg98}. We have in particular been interested in
examining to what extent the quasi-electron charge can be regarded as a
sharply defined quantum number. To study these problems we have
calculated charge fluctuations, as well as charge expectation values,
measured relative to the ground state.  We have done computations for a
Laughlin quasi-hole, and for quasi-electrons both with Laughlin's and
Jain's definition of the wave function. In all cases the quasi-particle
has been located at the center of a circular electron droplet.

For the quasi-hole charge we have found numerical results that are
consistent with the expected bulk value $-1/3$ (times the electron
charge) and charge fluctuations consistent with vanishing fluctuations
in the bulk. There are effects due to the finite size of the quasi-hole
and to the finite size of the droplet, but in an intermediate interval
the deviations from the expected values vanish within small statistical
errors. The computations have been done for systems with 50 and 100
electrons, and the results clearly confirm the conclusion that the
quasi-hole charge is a sharp quantum number.

For the Laughlin quasi-electron the calculations similarly reproduce
the expected bulk value of the charge, which is $1/3$ times the electron
charge. But for the charge fluctuations of this excitation we find
fluctuations which survive throughout the whole electron droplet. At an
absolute scale the values  are small, but we have shown that
they are significantly different from zero. We have studied systems
with 50, 100 and 200 electrons, and conclude that results obtained for
these system sizes do not unambiguously confirm the charge of the
Laughlin quasi-electron to be a sharp quantum number. The deviation from
zero which are found for the charge fluctuations may be a finite-size
effect, but the fluctuations then are sufficiently long range that even
a system of $200$ electrons is not sufficiently large to see this
clearly.

For a system with 50 electrons we have also studied the charge and
charge fluctuations of a Jain quasi-electron. In this case we again
reproduce expected results in the bulk, $1/3$ for the charge and
vanishing fluctuations. There are small deviations, but these are not
significant, within small statistical errors. We
also find that in this context there is almost no difference between the
results obtained for the Jain wave function projected onto the lowest
Landau level and the unprojected state. This confirms results presented
elsewhere in the literature, which show that the unprojected state is
almost entirely in the lowest Landau level.

As a further examination of the relation between sharpness of the charge
and a well defined statistics parameter, we have computed Berry phases
for the Jain quasi-electron. This gives a supplement to the previously
done calculations for the Laughlin quasi-electron \cite{Kjonsberg96}.
We have considered unprojected states for a system with 50 electrons,
and have extracted charge and statistics parameters in the usual way.
The charge parameter calculated in this way has a well defined bulk
value with only a small deviation from the expected value $1/3$. The
small deviation may be due the definition used for the wave function of
a quasi-electron translated to arbitrary position, but we have no firm
conclusion about this.

As the most interesting result of the Berry calculations, we found that
the statistics parameter of the Jain quasi-electron is much more well
behaved than the analogous quantity for the Laughlin quasi-electron.
With the convention used here, our results clearly indicate that
the value of this parameter (for large separations of the two
quasi-electrons) is $\nu=-1/3$, which is minus the statistics parameter
of the quasi-holes. However, the sign of the statistics parameter does
not agree with the expected one, and is not consistent with the value
$-2+1/3$ indicated by state counting of numerically determined energy
levels for quantum Hall states on a sphere. Thus, for neither of the
suggested wave functions, due to Laughlin and Jain, the expected
statistics parameter of the physical quasi-electrons seem to be
correctly reproduced.

\section*{Acknowledgments}

We are grateful to Hans Hansson for many helpful discussions and
useful suggestions throughout the work. In addition we would like to
acknowledge Geoffrey S. Canright for his help with making the
three-dimensional figures, as well as for several helpful
discussions, and Jan Myrheim for enlightening
comments. This work has received support from The Research Council
of Norway (Program for Supercomputing) through a grant of computing
time.

\appendix

\section{Numerical techniques} \label{numerikk}

This appendix reviews important aspects of the Metropolis algorithm
and Monte Carlo estimate.

In section \ref{s3} we give several examples of functions used to
generate electron configurations according to the Metropolis
algorithm, \eg Eq.~(\ref{prfir}). We will briefly review the
technical details of this method. Suppose then
that the desired (real) probability distribution is given by $p(z_1,\cdots,
z_N^*)$, and there is a given configuration
$\{z_{i,\alpha}\}_{i=1}^N$. To find the next configuration we loop
through the electrons, and for each $i$ we randomly choose a test
coordinate $z_{i}^{\rm t} = z_{i,\alpha} + \Delta z_i$ such that
$\Delta z_i$ and $-\Delta z_i$ are equally probable. We then compute
the ratio
\beq
f_i=   \frac{p(z_{1,\alpha},\cdots,z_i^{\rm t},\cdots, z_i^{*\rm t}, \cdots
z_{N,\alpha}^*)}{p(z_{1,\alpha},\cdots,z_{i\alpha},\cdots,
z_{i\alpha}^{*}, \cdots z_{N,\alpha}^*)}
\eeq
If the number $f_i$ is larger than a randomly generated number between
$0$ and $1$, then we accept the test coordinate and set
$z_{i,\alpha+1} = z_i^{\rm t}$. Otherwise $z_{i,\alpha+1} =
z_{i,\alpha}$. The procedure ensures that the configurations are
generated according to the desired probability distribution $p$
because the principle of detailed balance is satisfied by the
transition probabilities; the ratio of jumping from $z_a$ to $z_b$ or
from $z_b$ to $z_a$ equals the ratio $p(z_b)/p(z_a)$.

For the cases we have considered it is of special interest to notice
that an overall normalization factor in $p$ is irrelevant for the
Metropolis algorithm since the latter only considers ratios between
different value of $p$. In all our calculations of charge and charge
fluctuations we have benefited from this in the sense that we have
used probability distributions for which we did not know how to
analytically find the normalization factor. For instance the 
factor $I_0$ in
Eq.~(\ref{prfir}) is not known exactly. It was essential that we could
do this, because the computer time needed to obtain well converged
results highly depends on the choice of probability distribution. The
time is reduced when $p$ has a behaviour similar to the actual
integrand of the problem under consideration, because the standard
deviation of the Monte Carlo estimate is given by $\sqrt{V(g/p)/n}$,
with $g$ as the specific integrand, $V(g/p)$ as the variance of the
function $g/p$, and $n$
the number of Monte Carlo
steps.

\section{Relations between ${\widetilde Q}(R)$, 
${\widetilde F}(R_1,R_2)$ and
$Q_{\rm Le}(R)$, $F_{\rm Le}(R)$}  \label{relst}

In this appendix we will show how to derive the expressions in
Eqs.~(\ref{tilq},
\ref{tilf}).

We start by considering the relation between ${\widetilde
Q}(R)$ and $Q_{\rm Le}(R)$.
According to Eq.~(\ref{qno}) the desired charge expectation value can
be written
\be
Q_{\rm Le}(R) &=& \frac{N}{I_{\rm Le}}
\int_A {\rm d}^2\! z_1 \int {\rm d}^{2(N-1)}\! z \, |\psi_{\rm
Le}(z_1,\cdots, z_N^*)|^2 \\
&=& 2\pi  \int_0^R {\rm d}r \,r h(r),
\ee
where we have defined
\beq
h(r) = \frac{1}{2\pi} \int_0^{2\pi}{\rm d}\varphi \rho(z_1,z_1^*);
\;\;\;\;\;\;\; z_1 = r e^{i\varphi},
\eeq
with
\be
\rho(z_1,z_1^*) &=& N \frac{1}{I_{\rm Le}} e^{-|z_1|^2}
\partial_{z_1}\partial_{z_1^*} \\
&& \int {\rm d}^2 z_2 \cdots {\rm d}^2 z_N
e^{-\sum_{i=2}^N |z_i|^2} \prod_{k=2}^N (|z_k|^2-1) \prod_{i<j}|z_i -
z_j|^{2m}. \label{gfunuti}
\ee
It is important to notice the ranges of the counting variables in this
expression, which is obtained using integration by parts for all
coordinates that are integrated over the entire complex
plane. The function $\rho(z_1,z_1^*)$ is the exact single-electron
density. We now define the quantities
\beq
{\tilde \rho}(z_1,z_1^*) = N \frac{1}{I_{\rm Le}}
 \int {\rm d}^2 z_2 \cdots {\rm d}^2 z_N
e^{-\sum_{i=1}^N |z_i|^2} \prod_{k=1}^N (|z_k|^2-1) \prod_{i<j}|z_i -
z_j|^{2m}, \label{gti}
\eeq
\beq
{\tilde h}(r) = \frac{1}{2\pi} \int_0^{2\pi}{\rm d}\varphi {\tilde
\rho}(z_1,z_1^*)
\eeq
and
\beq
{\widetilde Q}_{\rm Le}(R) = 2\pi  \int_0^R {\rm d}r \,r {\tilde h}(r).
\eeq
The latter quantity is identical to the one defined in
Eq.~(\ref{tilQ}). Comparing Eqs.~(\ref{gfunuti}) and (\ref{gti}) we
observe that the relation between $\rho$ and $\tilde \rho$ can be
written
\beq
\rho(z_1,z_1^*) = e^{-|z_1|^2} \partial_{z_1}\partial_{z_1^*} \left(
e^{|z_1|^2} \frac{1}{(|z_1|^2-1)} {\tilde \rho}(z_1,z_1^*)\right).
\eeq
It is then straight forward to show that the radial functions $h$ and
$\tilde h$ are related by
\beq
h(r) = {\tilde h}(r) + \frac{1}{r} \frac{\rm d}{{\rm d}r} \left(
\frac{r^2}{r^2-1} {\tilde h}(r) +\frac{1}{4}r \frac{\rm d}{{\rm
d}r}\left(\frac{ {\tilde h}(r)}{ r^2-1}\right)\right).
\eeq
This implies that
\be
Q_{\rm Le}(R) &=&  2\pi  \int_0^R {\rm d}r \,r h(r) \\
&=& 2\pi  \int_0^R {\rm d}r \,r {\tilde h}(r) + 2\pi  \left(
\frac{R^2}{R^2-1} {\tilde h}(R) +\frac{1}{4}R \left(\frac{\rm d}{{\rm
d}r}\frac{ {\tilde h}(r)}{ (r^2-1)}\right)_{r=R}\right),
\ee
since the contribution from $r=0$ is zero. Notice that the appearant
singularity at $r=1$ is artificial since the factor $r^2-1$ in the
denominators is canceled by the same factor in the numerators; recall
the definition of ${\tilde h}(r)$. Using the fact that
\beq
2\pi {\tilde h}(R) = \frac{1}{R} \frac{{\rm d} {\widetilde Q}(R)}{{\rm
d}R}
\eeq
we finally obtain the advertised relation
\beq
Q_{\rm Le}(R) = {\widetilde Q}(R) + c_1(R) \frac{{\rm d}{\widetilde
Q}(R)}{{\rm d}R}
+ c_2(R)
\frac{{\rm d}^2{\widetilde Q}(R)}{{\rm d}R^2}, \label{tilqq}
\eeq
with
\be
c_1(R) = \frac{4R^4-7R^2+1}{4R(R^2-1)^2},  &&
c_2(R) = \frac{1}{4(R^2-1)}.  \label{c1c22}
\ee

To find the relation between $F_{\rm Le}(R)$ and 
${\widetilde F}(R_1,R_2)$
we perform a calculation that is similar in spirit, although
technically more complicated than the one above. First we define the
radial function
\beq
h_2(r_1,r_2) = \frac{1}{4\pi^2} \int_0^{2\pi}{\rm d}\varphi_1 {\rm
d}\varphi_2  \rho_2(z_1,z_2,z_1^*,z_2^*);
\;\;\;\;\;\;\; z_1 = r e^{i\varphi_1}, z_2 = r e^{i\varphi_2},
\eeq
with
\be
\rho_2(z_1,z_2,z_1^*,z_2^*) &=& N(N-1) \frac{1}{I_{\rm Le}}
e^{-|z_1|^2-|z_2|^2}
\partial_{z_1}\partial_{z_1^*} \partial_{z_2}\partial_{z_2^*} \\
&& \times \int {\rm d}^2 z_3 \cdots {\rm d}^2 z_N
e^{-\sum_{i=3}^N |z_i|^2} \prod_{k=3}^N (|z_k|^2-1) \prod_{i<j}|z_i -
z_j|^{2m} \label{gfunuti2}
\ee
as the true two-particle distribution function. We can then write
the desired function in Eq.~(\ref{fno}) as
\beq
F_{\rm Le}(R) = 4\pi^2 \int_0^R {\rm d}r_1 \,\int_0^R 
{\rm d}r_2 \, r_1 r_2
h_2(r_1,r_2).
\eeq
Let us now define the auxiliary quantities
\be
{\tilde \rho}_2(z_1,z_2,z_1^*,z_2^*)  &=& N(N-1) \frac{1}{I_{\rm
Le}} \\
&& \times \int {\rm d}^2 z_3 \cdots {\rm d}^2 z_N
 e^{-\sum_{i=1}^N |z_i|^2} \prod_{k=1}^N (|z_k|^2-1) \prod_{i<j}|z_i -
z_j|^{2m} ,\label{gfunuti2f}
\ee
\beq
{\tilde h}_2(r_1,r_2) = \frac{1}{4\pi^2} 
\int_0^{2\pi}{\rm d}\varphi_1 {\rm
d}\varphi_2  {\tilde \rho}_2(z_1,z_2,z_1^*,z_2^*),
\eeq
and
\beq
{\widetilde F}(R_1,R_2) = 4\pi^2 \int_0^{R_1} {\rm d}r_1\, r_1
\int_0^{R_2} {\rm d}r_2\, r_2 {\tilde h}_2(r_1,r_2).  \label{labf}
\eeq
Analogous to what we saw above, also now the relation between the
two-electron density $\rho_2$ and
the quantity ${\tilde \rho}_2$ is easily established, and it reads
\beq
\rho_2 = e^{-|z_1|^2-|z_2|^2} \partial_{z_1}\partial_{z_1^*}
\partial_{z_2}\partial_{z_2^*} \left( \frac{
e^{|z_1|^2+|z_2|^2}}{(|z_1|^2-1)(|z_2|^2-1)} {\tilde \rho}_2\right).
\eeq
A similar exact relation exists between the radial quantities
$h_2(r_1,r_2)$ and ${\tilde h}_2(r_1,r_2)$. We rewrite it
into a form suited for integration and find
\be
h_2(r_1,r_2) &=& {\tilde h}_2(r_1,r_2) \nonumber \\
&& + (r_2^2-1) \frac{1}{r_1} \frac{\partial}{\partial r_1} \left(
r_1^2 g_2\right)
 + (r_1^2-1)  \frac{1}{r_2}  \frac{\partial}{\partial r_2} \left(
r_2^2 g_2\right) \nonumber \\
&& + \frac{1}{4}(r_2^2-1)\frac{1}{r_1} \frac{\partial}{\partial r_1}
\left( r_1 \frac{\partial g_2}{\partial r_1}\right) +
\frac{1}{4}(r_1^2-1)\frac{1}{r_2}  \frac{\partial}{\partial r_2}
\left( r_2 \frac{\partial g_2}{\partial r_2}\right)\nonumber  \\
&& + \frac{1}{4} \frac{1}{r_1 r_2}\frac{\partial}{\partial r_1} \left(
r_1 \frac{\partial}{\partial r_1} \frac{\partial}{\partial r_2} (r_2^2
g_2)
\right)
+ \frac{1}{4} \frac{1}{r_1 r_2}\frac{\partial}{\partial r_2} \left(
r_2 \frac{\partial}{\partial r_1} \frac{\partial}{\partial r_2} (r_1^2
g_2)
\right) \nonumber \\
&& +  \frac{1}{r_1 r_2} \frac{\partial}{\partial r_1}
\frac{\partial}{\partial r_2} (r_1^2 r_2^2 g_2) \nonumber \\
&& + \frac{1}{16} \frac{1}{r_1 r_2} \frac{\partial}{\partial r_1}
r_1\frac{\partial}{\partial r_1} \frac{\partial}{\partial r_2}
r_2\frac{\partial}{\partial r_2} g_2. \label{ille}
\ee
We have in this expression used the notation
$g_2(r_1,r_2) = \frac{1}{(r_1^2-1)(r_2^2-1)}{\tilde h}_2(r_1,r_2)$,
and notice that $r_1$ and $r_2$ are treated symmetrically in
Eq.~(\ref{ille}).
In what proceeds we will not in detail evaluate the integral of
every term in this expression. That is a tedious job, and not very
informative. Instead, we will show an example that
is representative and hence give the idea for how to do the
other calculations. So let us consider the first term in
the second line of Eq.~(\ref{ille}), and integrate it the way we need
in order to find $F_{\rm Le}(R)$. This yields
\be
&&4\pi^2 \int_0^{R} {\rm d}r_1\, r_1
\int_0^{R} {\rm d}r_2\, r_2
(r_2^2-1) \frac{1}{r_1} \frac{\partial}{\partial r_1} \left(
r_1^2 g_2\right) \nonumber \\
&=& 4\pi^2 \int_0^{R} {\rm d}r_2\, r_2  (r_2^2-1)
\left[r_1^2 g_2(r_1,r_2) \right]_0^{r_1=R}  \nonumber \\
&=& 4\pi^2 \frac{R^2}{(R^2-1)} \int_0^{R} {\rm d}r_2\, r_2
{\tilde h}_2(R,r_2)  \nonumber \\
&=& 4\pi^2 \frac{R}{(R^2-1)}
\left(\frac{\partial {\widetilde F}(R_1,R_2)}
{\partial R_1}\right)_{R_1,R_2 = R}.
\ee
The last equality sign shows the relation between $\tilde h$ and
one of the derivatives of $\widetilde F$. This expression 
easily follows from
Eq.~(\ref{labf}). Other derivatives can be similarly
expressed, and we use the relations
\beq
\frac{\partial^2 {\widetilde F}(R_1,R_2)} {\partial R_1  R_2}
= 4\pi^2 R_1 R_2 {\tilde h}(R_1,R_2),
\eeq
\beq
\frac{\partial^2 {\widetilde F}(R_1,R_2)} {\partial R_1^2} 
= \frac{1}{R_1}
\frac{\partial {\widetilde F}(R_1,R_2)} {\partial R_1} + 4\pi^2 R_1
\int_0^{R_2} {\rm d}r_2\,r_2
\frac{\partial {\tilde h}(R_1,r_2)}{\partial R_1},
\eeq
along with trivial extensions to
$\frac{\partial^3 {\widetilde F}(R_1,R_2)} {\partial R_1^2  R_2} $
and
$\frac{\partial^4 {\widetilde F}(R_1,R_2)} {\partial R_1^2  R_2^2} $.
Performing the integrations of all terms in Eq.~(\ref{ille}) and
collecting together similar terms we end up with the expression in
Eq.~(\ref{tilf}), that is
\be
F_{\rm Le}(R) &=& {\widetilde F}(R,R) + c_1(R)\left( \frac{\p 
{\widetilde F}}{\p
R_1} + \frac{\p
{\widetilde F}}{\p R_2}\right)_R \nonumber \\
&& + c_2(R)\left( \frac{\p^2 {\widetilde F}}{\p R_1^2} + \frac{\p^2
{\widetilde F}}{\p R_2^2}\right)_R + c_1^2(R) \left( \frac{\p^2
{\widetilde F}}{\p R_1 R_2 }\right)_R \nonumber \\
&& + c_1(R) c_2(R) \left( \frac{\p^3 {\widetilde F}}{\p R_1^2 R_2} 
+ \frac{\p^3
{\widetilde F}}{\p R_1 R_2^2}\right)_R + c_2^2(R) \left( \frac{\p^4
{\widetilde F}}{\p R_1^2 R_2^2}\right)_R. \label{tilff}
\ee


\begin{thebibliography}{99}

\bibitem{exp} L. Saminadayer, D. C. Glattli, Y. Jin and B. Etienne,
{\em Observation of the $e/3$ fractionally charged Laughlin
quasiparticles}, Phys. Rev. Lett. {\bf 79} (1997)
2526. \newline R. de-Picciotto, M. Reznikov, M. Heiblum, 
V. Umansky, G. Bunin
and D. Mahalu, {\em Direct Observation of a Fractional Charge}, Nature
{\bf 389} (1997) 162.

\bibitem{Laughlin} R. B. Laughlin, {\em Anomalous Quantum Hall Effect:
An Incompressible Quantum Fluid with Fractionally Charged Excitations},
Phys. Rev. Lett. {\bf 50}
(1983) 1395.

\bibitem{Arovas84} D. Arovas, J. R. Schriffer and F. Wilczek, 
{\em Fractional
Statistics and the Quantum Hall Effect}, Phys. Rev. Lett. {\bf 53}
(1984) 722.

\bibitem{Jain} J. K. Jain, {\em Composite-Fermion Approach for the
Fractional Quantum Hall Effect}, Phys. Rev. Lett. {\bf 63} (1989) 199.

\bibitem{Kjonsberg96} H. Kj\o nsberg and J. M. Leinaas,
{\em On the anyon description of the Laughlin hole states},
Int. Jour. Mod. Phys. A {\bf 12} (1997) 1975.

\bibitem{Kjonsberg98} H. Kj\o nsberg and J. Myrheim,
{\em Numerical study of charge and statistics of Laughlin 
quasi-particles},
to appear in Int. Jour. Mod. Phys. A.


\bibitem{Kivelson85} S. Kivelson and M. Ro\v{c}ek,
{\em Consequences of gauge invariance for fractionally charged
quasi-particles},
Phys. Lett. {\bf 156B} (1985) 85.


\bibitem{Kivelson82} S. Kivelson and J. R. Schrieffer,
{\em Fractional charge, a sharp quantum observable},
Phys. Rev. B {\bf 25} (1982) 6447.

\bibitem{Rajaraman82} R. Rajaraman and J. S. Bell,
{\em On solitons with half integral charge},
Phys. Lett. {\bf 116B} (1982) 151.

\bibitem{Goldhaber91} A. S. Goldhaber and S. Kivelson,
{\em Local charge versus Aharonov-Bohm charge},
Phys. Lett. B {\bf 255} (1991) 445.

\bibitem{james} F. James, {\em Monte Carlo theory and practice},
Rep. Prog. Phys. {\bf 43} (1980) 1145.

\bibitem{laughlingp} R. B. Laughlin in {\em The Quantum Hall Effect}
eds. R. E. Prange and S. M. Girvin, Springer-Verlag, 1990.

\bibitem{halpmorf} R. Morf and B. I. Halperin, {\em Monte Carlo
evaluation of trial wave functions for the fractional quantized Hall
effect: Disk geometry}, Phys. Rev. B
{\bf 33} (1986) 2221.

\bibitem{hr}F. D. M. Haldane and E. H. Rezayi, {\em Finite-Size
Studies of the Incompressible State of the Fractionally 
Quantized Hall
Effect and its Excitations}, Phys. Rev. Lett. {\bf 54} (1985) 237.

\bibitem{japim} N. Trivedi and J. K. Jain, {\em Numerical Study of
Jastrow-Slater Trial States for the Fractional Quantum Hall Effect},
Mod. Phys. Lett. B {\bf 5}
(1991) 503.

\bibitem{girvjach} S. M. Girvin and T. Jach, {\em Formalism for the
Quantum Hall Effect: Hilbert space of analytic functions},
Phys. Rev. B {\bf 29} (1984) 5617.

\bibitem{num} W. H. Press, W. T. Vetterling, S. A. Teukolsky and
B. P. Flannery, {\em Numerical Recipes in C}, Cambridge University
Press, 1992.

\bibitem{berry} M. B. Berry, {\em Quantal phase factors accompanying
adiabatic changes}, Proc. R. Soc. Lond. A. {\bf 392} (1984) 45.

\bibitem{Aharonov59} Y. Aharonov and D. Bohm, {\em Significance of
electromagnetic potentials in the quantum theory}, Phys. Rev. {\bf
115} (1959) 485.

\bibitem{Hansson92} T. H. Hansson, J. M. Leinaas and J. Myrheim, {\em
Dimensional reduction in anyon systems}, Nucl. Phys. B 
{\bf 384} (1992) 559.

\bibitem{Haldane} F. D. M. Haldane, {\em ``Fractional statistics'' 
in arbitrary dimensions: A generalization of the Pauli principle}, 
Phys. Rev. Lett. {\bf 67} (1991) 937.  

\bibitem{Dev92} G. Dev and J.K. Jain, {\em Jastrow-Slater trial wave
functions for the fractional quantum Hall effect: Results for
few-particle systems}, Phys. Rev. B {\bf
45} 1223 (1992).

\bibitem{Canright94} M. D. Johnson and G. S. Canright, 
{\em Haldane fractional statistics in the fractional quantum 
Hall effect}, Phys. Rev. B 
{\bf 49} (1994) 2947.


\end{thebibliography}
\end{document}